\documentstyle[prl,aps,floats,epsf,epsfig]{revtex}

\begin{document}

\draft

\title{ The geometric order of stripes and Luttinger liquids.}

\author{J. Zaanen, O. Y. Osman, H. V. Kruis, and Z. Nussinov}
\address{Instituut Lorentz for Theoretical Physics, Leiden University\\
P.O.B. 9506, 2300 RA Leiden, The Netherlands}
\author{J. Tworzyd\l o}
\address{Institute of Theoretical Physics, Warsaw University, Hoza 69,
00-681 Warszawa, Poland}
\date{\today ; E-mail:jan@lorentz.leidenuniv.nl}

\twocolumn[

\widetext
\begin{@twocolumnfalse}

\maketitle

\begin{abstract}
It is argued that the electron stripes as found in correlated oxides have to do
with an unrecognized form of order. The manifestation of this order is the
robust property that the charge stripes are at the same time anti-phase boundaries
in the spin system. We demonstrate that the quantity which is ordering is
sublattice parity, referring to the geometric property of a bipartite lattice
that it can be subdivided in two sublattices in two different ways. Re-interpreting
standard results of one dimensional physics, we demonstrate that the same order
is responsible for the phenomenon of spin-charge separation in strongly interacting
one dimensional electron systems. In fact, the stripe phases can be seen from this
perspective as the precise generalization of the Luttinger liquid to higher dimensions.
Most of this paper is devoted to a detailed exposition of the mean-field theory of
sublattice parity order in 2+1 dimensions. Although the quantum-dynamics of the spin- and charge degrees
of freedom is fully taken into account, a perfect sublattice parity order is imposed.
Due to novel order-out-of-disorder physics, the sublattice parity order gives rise to
full stripe order at long wavelength. This adds further credibility to the notion that
stripes find their origin in the microscopic quantum fluctuations and it suggests a 
novel viewpoint on the relationship between stripes and high Tc superconductivity.
\end{abstract}

\vspace{0.5cm}

\narrowtext

\end{@twocolumnfalse}
]

\section{Introduction.}

This rather long paper is devoted entirely to a possible answer to the
question {\em what are stripes ?} This might sound odd since the electron
stripes as found in correlated oxides have grown into a popular subject.
Of course there is a popular answer to this question: the `rivers of
charge' separated by (quasi) insulating domains (Zaanen, 1999 (2)). However, this answer
is too intuitive and lacks the precision needed in the context
which matters most: are stripes central to the problem of high Tc
superconductivity, or do they represent a red herring, a real but
irrelevant side effect?

In the light of the history of the subject, nobody can afford 
to be convinced of anything related to high Tc superconductivity. Nevertheless, one
might want to stick to the most general principles, namely those of
symmetry. The high Tc riddle has to do with the highly anomalous macroscopic properties
of the electron system of the cuprates. Hence, the question is on
the long wavelength properties of  the system and here one is helped
by general symmetry-based considerations, of the kind called by Laughlin `competing orders' (Laughlin, 1998).
There appear to be  two possibilities:
either the fixed point is adiabatically connected to the BCS 
superconductor or the fixed point is a different one. In the first
case, one has to explain why the fixed point is approached by a highly
anomalous cross-over and this suggests the proximity of a different
fixed point. In the second case, high Tc superconductivity is
separated from a conventional superconductor by a non-adiabatic
boundary and this implies that the symmetries of both states are
different. The high Tc superconductor is surely a Meissner phase, and
the difference in symmetry is somewhere else. It is apparently
hard to detect by experiment and therefore it is called `hidden order'.
Also if it is a mere crossover behavior, the nearby competing state 
is apparently equally hard to detect experimentally and one might
want to consider it as a variation on the hidden order theme.

Let us consider stripes from this perspective. Are they candidates
for the hidden order? It is a popular thought  that stripes are
just charge order (or charge density wave order, or just
an interesting Wigner crystal: it is all the same). The charge density of the
electrons breaks translational invariance. Given the stability 
of the Mott-insulator it is not remarkable that this charge density
wave is such that the hole poor regions become charge commensurate
with the crystal lattice. Hence, these turn into magnetic domains
and the subsequent breaking of spin-rotational invariance is then
considered to be a parasitic effect.

Can this charge order be the hidden order? Despite some brave
suggestions (Castellani, Grilli and di Castro, 1995), 
this appears as highly  unlikely. Translational
symmetry breaking is easy to detect and it is not seen
in experiment in the
vicinity of optimal doping in the best superconductors. Of course, this could have been missed
in the scattering experiments for technical  reasons. However,
it is by now well established that nodal fermions
exist also at  lower dopings (Orgad {\em et al.}, 2001) and this fact is very hard to  reconcile
with a significantly developed stripe charge order. 

An alternatively candidate could be the stripe magnetism. In fact, the incommensurate
magnetic fluctuations (often associated with the stripe magnetism) behave in a
manner which is reminiscent of the competing order hypothesis. The case has been made 
that they demonstrate the `quantum critical' scaling behavior associated with the proximity to a
continuous quantum phase transition (Aeppli {\em et al.}, 1997). However, to arrive at a more
detailed interpretation one has to invoke a strong spin-charge separation (Chubukov and Sachdev, 1993; Sachdev, 2000;
Zaanen, 1999(1); van Duin and Zaanen, 2000). The
magnetism goes its own way (presumably described by a quantum non-linear sigma
model) regardless of the charge dynamics. This hypothesis is directly violated
by experiment. It is clear that the gap seen in the spectrum of incommensurate
spin fluctuations opens up right at the superconducting transition (Dai {\em et al.}, 1999; Lee {\em et al.}, 2000).

Hence, it can be argued successfully that neither the charge order, nor the
magnetic order associated with the stripes can be held responsible for the
long wavelength anomalies of the high Tc phenomenon. The reason is that
the empirical consequences of these conventional orders are too well
understood and too easy to measure. 

The main aim of this paper is to illustrate the idea that  the above might be an incomplete
characterization of the symmetry structure of the stripe phase. Stripe order implies
that yet another symmetry is spontaneously broken and this symmetry structure is of 
a most unconventional kind.  On a heuristic level it is
widely recognized that something unusual is going on and  this is
called `topological doping' or `anti-phase
boundarieness' (Zaanen and Gunnarsson, 1989; Kivelson and Emery, 1996; 
Zaanen, Horbach and van Saarloos, 1996; Zaanen, 1998; Pryadko {\em et al.}, 1999). 
All available experimental information supports
the notion that the charge stripes are at the same time anti-phase
domain walls in the anti-ferromagnet (Zaanen, 2000(2)). This anti-phase boundarieness
is robust. It is not only there for static stripes (Tranquada {\em et al.}, 1995, 1999; 
Emery, Kivelson and Tranquada, 1999); the `dynamical
stripes' are also defined through this anti-phase boundarieness. Most
of the information on the  latter comes from the spin-fluctuations
as measured by inelastic neutron scattering. The interpretation of
these in terms of stripes rests in first instance on the 
characteristic wave-vectors of these fluctuations and their dependence
on doping (Aeppli {\em et al}, 1997; Mook and Dogan, 1999; Mook {\em et al.} 2000;
Dai {\em et al.} 1999; Yamada {\em et al.}, 1998; Lee {\em et al.}, 2000).
 This assumes anti-phase boundarieness which extents up
to large energy scales. This anti-phase boundarieness is also a
common denominator in many theoretical works addressing the microscopy
of the stripe phase, ranging from the early mean-field work to
the sophisticated recent exact diagonalization  studies (Zaanen and Gunnarsson,
1989; Zaanen, 1998; White and Scalapino, 1998; Morais-Smith {\em et al.} 1998; Pryadko {\em et al.}, 1999;
Fleck {\em et al.}, 2000; Martin {\em et al.}, 2000; Stojkovic {\em et al.}, 2000;
Tchernyshyov and Pryadko, 2000). Here we will
largely ignore stripe microscopy and instead try to contribute to
the understanding of the long wavelength dynamics from a phenomenological
perspective. We want to suggest that this anti-phase boundarieness
is a manifestation of {\em spontaneous symmetry breaking}. 

Stripes are not completely on their own in this regard:
 {\em the same symmetry principle is behind the 
phenomenon of spin-charge separation in the Luttinger liquids
of one dimensional physics.} In fact, symmetry-wise
stripes should be understood as the unique realization of
spinful Luttinger liquids in 2+1D. 

The expert in one dimensional physics might find this puzzling:
why should spin-charge separation have anything to do with 
spontaneous symmetry breaking? This will be discussed in
some detail in section II. We start out with well-known, 
mathematical  results in 1+1D physics (Ogata and Shiba, 1990) to find
that these can be reformulated in the language of order
parameter correlators after identifying the degree of
freedom which is carrying the order. This degree of
freedom is quite simple but counter-intuitive:
for lattice problems it is {\em sublattice parity} (Zaanen and Nussinov, 2001).

Hence, we claim that the Luttinger liquid is characterized
by an order parameter. Sublattice parity is an Ising ($Z_2$) degree
of freedom and true long range order can therefore exist at zero temperature
in 1+1D.  In the order parameter formulation it becomes trivial
to generalize spin-charge separation to higher dimensions and in 2+1D this
turns into the anti-phase boundarieness of the stripes. 

As will be discussed in section II, one might want to view this sublattice parity
order theory  as a parametrization in terms of auxiliary degrees of freedom
of a more fundamental {\em geometric} structure. `Geometric' is used here in
the same sense as in the Einstein theory of gravity. The geometry of embedding space is
different for an external observer and the internal observer, in the present case the 
experimentalist and the {\em spin system}, respectively.
However, this geometry is very simple, at least as compared to that of fundamental space-time.
The spin system is a quantum antiferromagnet and the only property of embedding space which
matters is the bipartiteness of the effective lattice seen by the spins. Full order in
the sublattice parity language implies that the true lattice seen  by the spin system is
bipartite everywhere and thereby geometrically unfrustrated. In one dimension it is always
possible to divide a lattice in two sublattices and we suspect that this offers a special
protection to the spin-charge separation phenomenon. However, in higher dimensions bipartiteness
is not automatic. In the order parameter language, topological excitations of sublattice 
bipartiteness order can be identified in 2+1D which cannot exist in 1+1D. These correspond
in the geometrical language with curvature events, equivalent to essential  frustrations
in the spin system (Zaanen and Nussinov, 2001). Although it demonstrates that in 2+1D sublattice parity order is not generic,
the precise nature of the {\em disorder} theory is at present very poorly understood -- it is a
most unusual structure. We suspect that this structure has something to do with the mysteries of
high Tc superconductivity, and in the final section we will discuss some work in progress to 
illustrate the problem.

The remainder of this paper (Sections III-VII) might be considered as a review on the part of the problem we understand
fairly well. It summarizes a large amount of work carried out during the last 7 years  by our
group in Leiden (Zaanen, Horbach and van Saarloos; 1996; Zaanen, 1996; Eskes {\em et al.}, 1996;
Zaanen and van Saarloos, 1997; van Duin and Zaanen, 1998; Eskes {\em et al.}, 1998;
Zaanen, Osman and van Saarloos, 1998; Tworzydlo {\em et al.}, 1999; Zaanen, 2000) 
When we started this pursuit we were  convinced that we were addressing rather unrelated
parts of the physics.  Amusingly, as a lucky circumstance we just looked at all the bits
and pieces needed to arrive at a  synthesis which we recognized only quite recently. 
It starts out with a single assumption, defining an important limiting case of the general problem: in
sections III-VII we present the {\em mean-field} theory of sublattice parity order in 2+1D. All we require
is that sublattice parity order is {\em perfectly} obeyed in 2+1D. We even do not allow for local 
violations of this order, let alone the global violations as discussed in the previous paragraph.
At first sight this amounts to a quite mild, partial constraint on the dynamics. The charges have still the 
freedom to delocalize and the spin system is a highly quantum-mechanical entity as well. The sublattice
parity order just amounts to the requirement that the charges have to form connected lines, like Ising
domain walls. A rather lively `toy universe' emerges and we refer to section III for an overview.
Nevertheless, the  sublattice parity order exerts an unexpected dominance. {\em The mere presence of perfect sublattice
parity order forces under all circumstances the charge and the spin to order as well.}
This is due to a rather counter-intuitive order-out-of-disorder
mechanism: the more severe the microscopic quantum fluctuations, the more robust the long range order at
large scales. Although this story does not solve any of the pressing problems in the high $Tc$ context we do
find it interesting. Albeit being the wrong limit, it corresponds with a reference point for the construction
of a more complete theory which might relate to high Tc superconductivity and we find the complexity of this
mean-field theory fascinating. On a practical level, it adds a counter-intuitive  meaning to the notion that 
stripe long range order can originate in the microscopic quantum fluctuations of spins and holes.

\section{Sublattice parity order.}

In this section we will not come up with anything truely new. It is just a recollection of some
well known facts of one dimensional physics. However, our consideration is focused on exposing
the bare essence of what is meant by charge-spin separation. This will turn out to be
a remarkably simple principle which can be trivially imposed in higher dimensions. A more extensive
discussion will be published elsewhere (Kruis, 2001). 

Charge-spin separation refers to the general property of one dimensional electron systems that the electron is
an unstable particle while at the same time the real propagating excitations of the system carry fractions of 
the charge of the electron and the spin separately (Anderson, 1997). This property can be deduced in various ways. It was
first found in the bosonization framework and later confirmed in the exact Bethe Ansatz solutions (Voit, 1994). It is
clearly related to a topological structure. Remarkably,  this has to do with a {\em kink} topology, the
topological structure associated with an Ising type field. This should not be considered as self-evident;
the manifest symmetry of the problem is $SU(2) \times U(1)$ (the spin- and charge global symmetries), and
why should there be a $Z_2$ Ising
 topological structure at work? Exactly the same problem is encountered in the
stripe context in 2+1 dimensions. For every theoretical purpose, the spin system in the cuprates is $SU(2)$
invariant. Why is it so that stripes are like Ising domain walls?

\begin{figure}[t]
\centering
\vspace{0.5cm}
\epsfig{figure=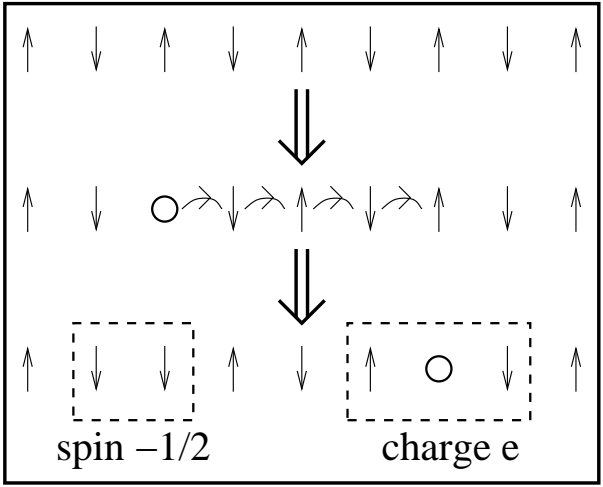,width=8cm}   
\vspace{0.5cm}
\caption{Cartoon picture of the mechanism of spin-charge separation in one dimension. A hole is injected in an antiferromagnetic $S=1/2$ spin-chain (top). When the hole moves to the right the spin moves backward (middle). The result of this kinematical
process is that a spin-domain wall carrying a $S=1/2$ quantum (spinon) is left at the origin while the hole is bound
to a spin anti-domain wall, and this composite only carries charge (bottom).}
\label{1dcartoon}
\end{figure}

In the one dimensional context one gets a first glimpse of the answer by considering a very simple
and well known example. Consider a 1D antiferromagnetic chain of Ising spins and remove a single spin 
somewhere as depicted in
 Fig. (\ref{1dcartoon}). Now consider what happens when the spin vacancy or hole moves to the left. The spin neighboring
the hole moves in the opposite direction and after a couple of hops one finds the hole surrounded by
anti-parallel spins while two parallel spins reside at the origin. Another way of calling this fact is that
the electron has split apart in a pure $S_z = 1/2$ excitation (the spin domain wall at the origin, or `spinon')
and a pure charge excitation (the `holon') because the spin of the original electron and that of the spin domain wall
carried around by the hole (the anti-parallel spins surrounding the hole) compensate. 
One directly infers that this has nothing to do with the Ising symmetry of the spins.
All what matters is that the spins have anti-ferromagnetic interactions while the motion of the vacancy
is accompanied by the backward motion of the spin. The net effect of attaching the domain wall to the hole
is that the backward moving spin ends up in the same, predominantly antiferromagnetic orientation relative
to its neighbors after the hole  has passed. Insisting that the spins surrounding the hole are parallel would imply
that every move of the hole would shake-off a spinon and this would be a very costly affair.

The implication is that this simple effect  also applies to Heisenberg
spins. Imagining, for
whatever reason, that these spins could be made to order in a perfect antiferromagnet, we could have as well
chosen to orient the spins along, say, the $x$ direction instead of the $z$ direction and the hole would
still have been surrounded by anti-parallel spins. It is a less
trivial matter to see that this effect is robust against
quantum spin fluctuations. This has been proven rigorously to be the case, as we will discuss later, but
it can already be inferred using  a simple continuation argument for the single hole case. Although the
microscopic spin fluctuations for the $S=1/2$ case are severe, at long distances the classical N\'eel
state is closely approached as signaled by the algebraic spin-correlations (Haldane, 1981). 
 The hole domain wall is topological so that it exerts its influence at infinity, and at infinity 
the spins are closely approaching the classical limit. Consider the
following Gedankenexperiment. Start out with the Ising spin chain and inject a single, completely delocalized
Ising-holon which is, however, constrained to move in between  two charge-potential barriers placed far apart. By measuring the
spin correlator at two points outside and at opposite sides of these barriers one easily infers that a 
holon has to be around because an up spin resides in the region to the right on the down spin sublattice of the region
to the left and vice versa. Switch
on subsequently the XY terms in the spin Hamiltonian. Upon reaching the Heisenberg point the spin 
correlations change from true long range order to algebraic order. It has to be that the kink is
still around because it violates the algebraic correlations in the same way as it violates the
true long range order of the Ising spins -- a single Ising kink in one dimension suffices to cause true disorder.

The nature of the Ising field
supporting the kink remains to be clarified -- from the previous
discussion it is clear that this Ising field is unrelated to the spins themselves. Although implicit to considerations
of the kind discussed in the previous paragraph (Schulz, 1993), it seems that this field
has not been explicitly identified before: it is sublattice parity. Sublattice parity is defined as follows.
Subdivide the lattice in $A$ and $B$ sublattices. Take an arbitrary reference point and start counting
with either $A-B-A-B- \cdots$ or $B-A-B-A- \cdots$ and call the two possibilities $1$ and $-1$, respectively.
This is an Ising variable, except for the subtlety that the global degeneracy is a gauge degeneracy 
associated with the arbitrariness of the reference site: one could have as well started the counting from a 
neighboring site. The holon is a hole bound to a domain wall in the sublattice parity: the A sublattice
changes into the B sublattice upon traversing the hole and vice versa. The simple kinematical
effect discussed in the previous paragraph translates into a {\em geometrical} principle governing the
collective dynamics. The only property of the embedding space which matters for a lattice quantum
antiferromagnet is its bipartiteness. The charge of the electron `curves' the space as seen by
the spin system, because it flips the parity of this bipartiteness. 

\begin{figure}[t]
\centering
\vspace{0.5cm}
\epsfig{figure=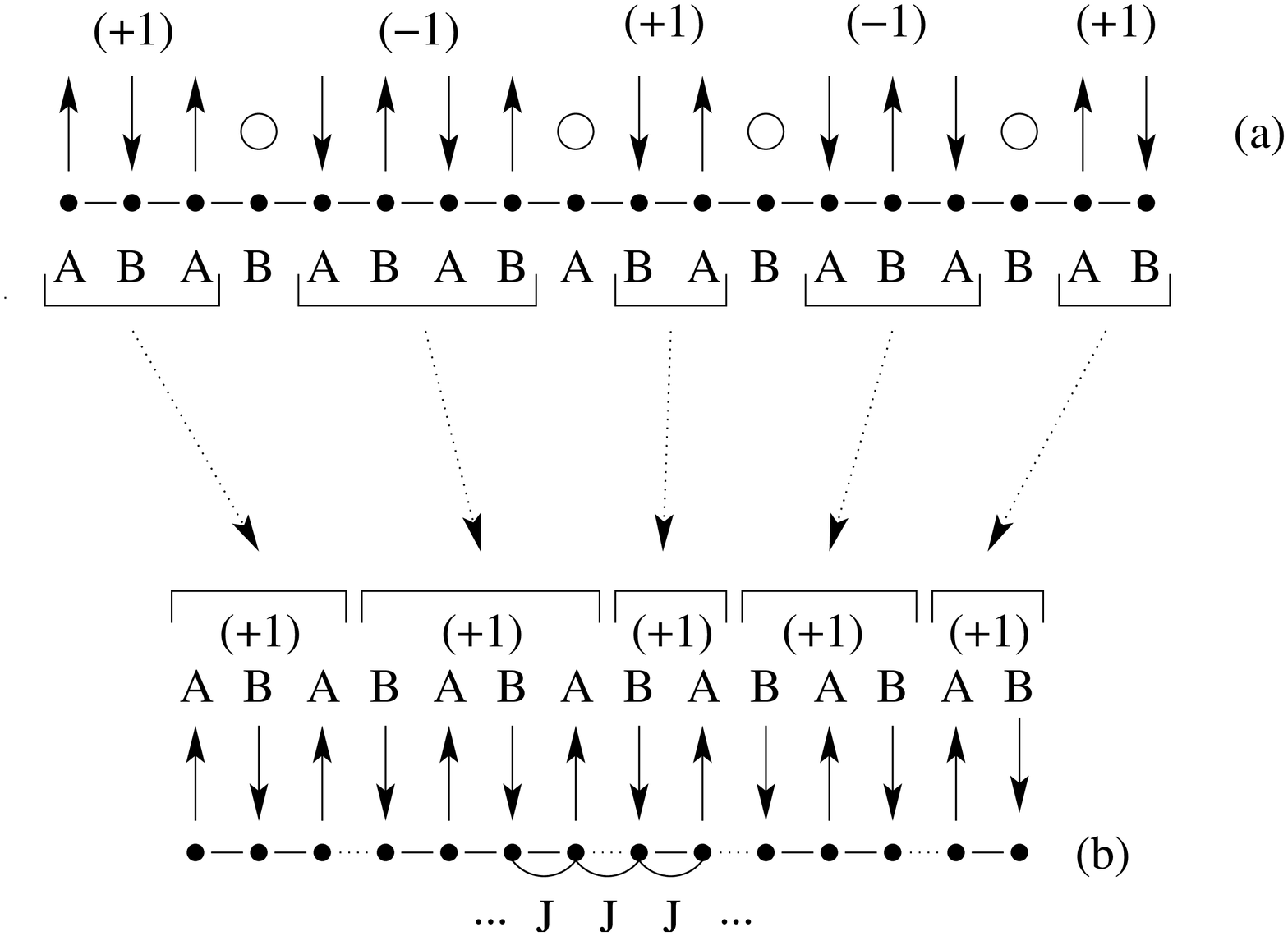,width=8cm}   
\vspace{0.5cm}
\caption{(a) According to the Bethe-Ansatz in the large $U$ limit (Ogata and Shiba, 1990) one can first
distribute the charges of the electrons and these charge configurations carry an amplitude equal to
that of an equivalent system of hard-core bosons, regardless the nature of the spin configurations. The
spin dynamics is that of a pure Heisenberg spin chain, where however the chain is a different one than
the original Hubbard chain. (b) This `squeezed chain'  is obtained by removing the holes and the sites
where the holes reside, substituting an antiferromagnetic exchange bond for these removed sites. This
squeezing operation is precisely equivalent to attaching sublattice parity flips to the holes in the
original Hubbard chain.}
\label{squeeze}
\end{figure}

A much better job can be done, and this was already accomplished some time ago by 
Ogata and Shiba (Ogata and Shiba, 1990; Anderson, 1997). They
observed that the Bethe-Ansatz ground state wave function of the Hubbard chain has the remarkable property 
in the large $U$ limit that it factorizes in a spin and a charge part. Consider a Hilbert space spanned
by Ising spin configurations and holes. In the large $U$ limit, as a first step one assigns the position
of the holes and according to the Bethe-Ansatz analysis 
this configuration has an amplitude equal to the amplitude of an equivalent system of hard core bosons,
regardless the configurations of the spins. Keeping the holes fixed, the spin-amplitudes follow from a
pure Heisenberg spin system which lives on a chain which is geometrically altered. This is the `squeezed'
Heisenberg chain, derived from the original Hubbard chain by removing the holes, and with the holes the
sites where they reside, reinserting an antiferromagnetic exchange coupling $J$ between the sites 
which were on both sides of the hole in the original chain, see Fig. \ref{squeeze}. 
The true spin dynamics is the one of the pure spin system living on the squeezed chain, thereby
explaining why spin and charge go their independent ways. As can be inferred from Fig.
\ref{squeeze},
this squeezing operation is precisely equivalent to  attaching sublattice parity flips to the holes in
the Hubbard lattice, which is just decoding the true space in which the spin system lives in the
fake space of experimentalists observing the full Hubbard chain. Fundamentally, the principle underlying
spin-charge separation seems to be best understood as a geometric principle -- the space in which
the spin dynamics lives is different from the Hubbard chain. However, the geometry involved is
exceedingly simple (bipartiteness) and it is trivially parametrized in terms of an order parameter
theory. This is an Ising theory. Every spin domain in between two holons is represented by an
Ising variable taking a value $\pm 1$ coding the value of the sublattice parity.
Every hole charge is  an antiferromagnetic exchange interaction between
these Ising spins. In the large $U$ limit there are apparently no fluctuations at zero temperature
and a perfect Ising order is established: $\cdots (+1) -0- (-1) -0- (+1) -0- (-1) -0- \cdots$.  
Spin-charge separation is just an Ising antiferromagnet!

The examples discussed in the above both refer to rather specific situations (strong coupling Hubbard
models), and it is a-priori unclear if the Ising order discussed in the previous paragraph is general.
To get anywhere, what is needed are order parameter correlators which can be explicitly calculated.
These are necessarily of an unconventional kind: sublattice parity order can only be measured using
the spin system because it parametrizes a geometric property of the space in which the spins live. 
 The strategy
is as follows: by inspecting the strong coupling limit a non-local correlator can be deduced which removes
the sublattice flips attached to the moving charges. This correlator thereby measures the true spin-correlations
living in the squeezed embedding space-time, isolating its spin-only character. By inspecting this correlator
one can indirectly infer the presence or absence of the sublattice parity order.

Define the staggered magnetization as usual as $\vec{M}_x = (-1)^x \vec{S}_x$ ($x$ is the site index). Define
also the charge operator $n_x$ taking the values $0, 1, 2$ for an empty, singly, and doubly occupied site.
Take an arbitrary point $x_0$ on the chain and define the following non-local (topological spin) operator
(Zaanen and van Saarloos, 1997),
\begin{eqnarray}
\vec{T}_{x_0, x} & = & e^{i \pi \sum_{y=x_0}^x ( 1 - n_y ) } \vec{M}_x \nonumber \\
\vec{T} (x_0, x) & = & e^{i \pi \int_{x_0}^x dy \left[ 1 - n(y)\right] } \vec{M} (x)
\label{topop}
\end{eqnarray}
where the second line is the corresponding expression in the continuum limit. Now consider the correlator
\begin{equation}
O_{top} ( | x - x'|, x_0 ) =
 \langle \Psi |\vec{T}_{x_0, x} \vec{T}_{x_0, x'} | \Psi \rangle
\label{topcor}
\end{equation}
the meaning of the `charge string 
operator' $\exp \left[i \pi \sum_{y=x_0}^x 
( 1 - n_y ) \right]$ is that it adds a
minus sign every time that a hole is passed on the trajectory $x_0 - x$. One infers immediately that this
charge string operator removes the flips in the sublattice parity attached to the holes. Instead of the
antiferromagnetic sublattice parity order seen by the standard spin correlator $ \cdots (+1) -0- (-1) -0- (+1) \cdots $
the topological correlator sees a ferromagnetic sublattice order $\cdots (+1) -0- (+1) -0- (+1) \cdots$ because of the
additional sign picked up every time a hole is passed. The topological correlator is easily evaluated in strong
coupling and instead of the usual result ($K_{\sigma}$ and $K_{\rho}$ are the spin  and the charge stiffness,
while $\varepsilon = 2k_F - \pi/a = \pi / n_h$, where $n_h$ is the hole density),
\begin{eqnarray}
O_{spin} ( | x - x'| ) & = & \langle \Psi |\vec{M}_{ x}
 \vec{M}_{ x'} | \Psi \rangle \nonumber \\
                       & = & B_{\sigma}{ { \cos (\varepsilon x)}
 \over { | x - x'|^{K_{\sigma}+K_{\rho} } } }
\label{spincor}
\end{eqnarray}
It is found that
\begin{equation}
O_{top} ( | x - x'|, x_0 ) = B_{\sigma} 
 \frac{1- n_h}{ | x - x'|^{K_{\sigma}}} 
\label{topcor1}
\end{equation}
Except for the `dilution factor' $1 - n_h$ Eq. (\ref{topcor1}) coincides with the correlator of a pure spin chain!
The direct  spin correlations as measured by Eq. (\ref{spincor}) decay more rapidly because they are sensitive 
to the antiphase-boundarieness attached to the charges which are  disorder events for the spins. Since the charges
exhibit algebraic order this adds only a simple additional  algebraic decay to the spin correlations
$\sim | x - x' |^{-K_{\rho}}$. The charge-string operator removes the anti-phase boundaries from the spin sector
and $O_{top}$ measures the physical spin correlations as they exist in the squeezed chain.

By analyzing the standard lore of one dimensional physics (Voit, 1994) we have arrived at the conclusion that
at least in strongly interacting lattice systems in 1+1D
charge-spin separation is controlled by a hidden order parameter. This order parameter structure
is in turn breaking an Ising symmetry and should therefore be regarded as rather robust because it is 
protected by a mass-gap -- it is the only true long range order which can exist
in the one dimensional Luttinger liquid!
Let us now proceed in a  phenomenological fashion. States of matter can only be rigorously defined through
their symmetry structure and we define the Luttinger liquid in arbitrary number of dimensions as states of matter
which exhibit the same  sublattice parity order as the Luttinger liquid in one dimension. This we find semantically 
more precise than the widespread habit of attempting to define the Luttinger liquid in higher dimensions through the
nature of the excitations (Anderson, 1997). The attentive
reader should already have realized the inescapable conclusion: static stripes are the genuine generalizations
of the Luttinger liquid to higher dimensions!

Let us consider the generalization of the sublattice parity order to higher dimensions in  more 
detail. Curiously, it appears to be not possible in all cases. Unlike
the one dimensional scenario, not all 
higher dimensional lattices can be partitioned into two sublattices. It is far from clear how to
construct stripe states on tripartite, etecetera, lattices and we suspect some interesting connections
between the geometric frustration of the spin system and the topological interplay of charge and spin.
It is likely not coincidental that the stripes  have only been found up to now in crystals characterized 
by bipartite lattices. 
\begin{figure}[t]
\centering
\vspace{0.5cm}
\epsfig{figure=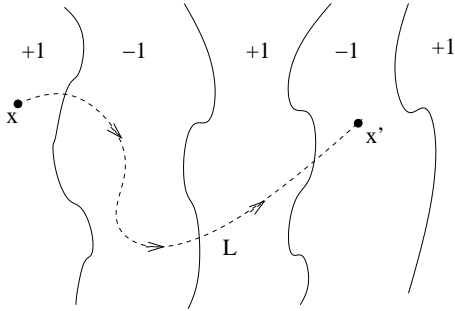,width=6cm}   
\vspace{0.5cm}
\caption{In two dimensions, the charge string correlator
corresponds with a line integral and the correlations also depend on the
length of the path $L$, see text.}
\label{cor2d}
\end{figure}

On square (cuprates, nickelates) or cubic (managanites) lattices, we can define a sublattice parity
Ising variable just as in the one dimensional case. The charge is  attached to domain
walls in the sublattice parity and Ising domain walls are lines in 2D, sheets in 3D, etcetera:
the stripes. As in the one dimensional case, by comparing the topological- and direct spin 
correlators one can learn if the sublattice parity order is established. Stronger, in space dimensions 
larger than one the topological correlator becomes more powerful. One can establish the sublattice
parity order by inspecting only the topological correlator. Consider the generalization of Eq.(\ref{topcor}) to higher
dimensions, 
\begin{equation}
O_{top} ( |\vec{x} - \vec{x}'|, L ) = \langle \Psi | \vec{M} (\vec{x}) 
e^{i \pi \int_{L,x}^{x'} dy \left[ 1 - n(y)\right] } 
\vec{M} (\vec{x}')| \Psi \rangle
\label{2Dtop}
\end{equation}
the charge-string operator corresponds with a line integral and the correlations no longer depend only on
the distance between the endpoints $\vec{x}-\vec{x}'$ but also on the length of the path $L$ of the path over which
the integral is evaluated (see Fig. \ref{cor2d}). True long range sublattice parity is established if the following
condition is satisfied,
\begin{equation}
\lim_{ |\vec{x} - \vec{x}'|, L \rightarrow \infty } O_{top} ( |\vec{x} - \vec{x'}|, L ) \rightarrow
G (| \vec{x} - \vec{x'} | )
\label{otoplim}
\end{equation}
where $G$ is only a function of the distance between the end points. According to the present understanding
of the stripe phenomenon, this condition should  be satisfied in the static stripes of cuprates and
nickelates. In fact, Zachar's recent analysis (Zachar, 2000) on the nature of the stripe disorder as driven by quenched
disorder can be taken as a direct evidence that the condition Eq. (\ref{otoplim}) is satisfied in the
static stripe phases of the cuprates.

\section{The mean-field theory of sublattice parity order.}

In the theory of order a most useful theoretical device is the limit where the order is perfect.
This perfect order is barely ever realized. However, as long as the violations of the order are
only local, the physics at long distances is qualitatively identical to that of the fully ordered
case. For most of the remainder of this paper we focus on the consequences of perfect sublattice order
in 2+1 dimensions. Hence, we impose that sublattice parity order is perfect which is equivalent to the statement 
that the internal space after the Ogata-Shiba squeeze is a bipartite 2D lattice, which is in turn equivalent
to the statement that the charges are attached to Ising domain walls in the sense that they
have to form {\em connected} $d-1$ dimensional manifolds in the embedding space with $d$ space dimensions.

The shear length of this paper makes already the point: this mean-field theory
of sublattice order is quite an interesting affair. The reason is that 
the presence of  sublattice parity long range order leaves much room for other physics to happen, 
and the 2+1 dimensional case is in this regard more interesting than the Luttinger liquid of 1+1D.
As introduction to the remainder of this paper, let us introduce the several physics problems 
which emerge after imposing the sublattice parity order:\\

(a) In principle, the $d-1$ dimensional stripe manifolds can have arbitrary shapes
and the question arises what happens in such
a system of interacting `branes'. Specifically, in 2+1 dimensions stripes are
lines on the time-slice and allowing for the fluctuations this represents a 
problem of {\em interacting quantum strings}, a string quantum fluid in 2+1
dimensions. This has been studied in great detail (Morais-Smith {\em et al}, 1998;
Dimashko {\em et al.}, 1999; Hasselmann {\em et al.}, 1999; Chernyshev {\em et al.}, 2000; 
Tchernyshyov and Pryadko, 2000), especially so in Leiden (Zaanen, Horbach and van Saarloos,
1996; Zaanen {\em et al.}, 1996; Eskes {\em et al.}, 1996, 1998; Zaanen, Osman and 
van Saarloos, 1998; Zaanen, 2000). 
and we have acquired quite some insight in the nature of this problem. We 
were surprised several times. A first surprise is that the mere presence of
an underlying lattice makes the problem of a single fluctuating stripe quite
tractable. In fact, as compared to the continuum string theory of high energy
physics these lattice strings are rather uninteresting objects: they either
pin to the lattice or they renormalize in free strings, as will be discussed
in Section IV. In addition, some general statements can be made about the system
of interacting strings. Given the condition of complete connectedness, a case
can be made that the system of strings in the presence of any interaction will
always order (Zaanen, 2000; Mukhin, van Saarloos and Zaanen, 2001). 
Given the results for the single string, the problem is obvious:
a single free string already exhibits algebraic translational order, and it is
obvious that the tendencies towards full order will be strong in a system of
such strings. A particularly subtle case is the one where the strings only
communicate via a non-crossing (hard-core) condition. This can be seen as
as the decompactified (to 2+1 D) version of the gas of non-interacting spinless
fermions in 1+1D and the argument will be reviewed in Section V, demonstrating
that even this string gas eventually orders, due to an order-out-of-disorder
mechanism.\\

(b)  Although the spin system is at least globally unfrustrated
in the presence of sublattice parity order (the `squeezed' lattice is bipartite) 
the quantum-spin physics in the 2+1D case is still quite rich. The reason is
that the stripes `slice' the spin system in 1+1 D ladders, and the 
quantum-magnetism of the static stripe system can be discussed in terms of 
coupled ladders (Tworzydlo {\em et al.}, 1999; Sachdev, 2000), 
as will be reviewed in Section VI. A next problem is, what
happens when the stripes are themselves strongly quantum fluctuating? In 
Section VII, we will present the results of a quantum Monte-Carlo simulation on a 
model which both incorporates the fluctuating stripes of Sections IV and V 
and the quantum spin dynamics of section VI (Osman, 2000). This has not been published before and it
might be considered as the most sophisticated stripe model studied up to now.
Besides illustrating vividly the physics discussed in the previous sections,
it also adds a next  piece of order-out-of-disorder physics: if the 
microscopic quantum-stripe fluctuations are sufficiently strong the spin system 
re-invents the classical N\'eel order, even if the spins of the fully static 
stripe system are  quantum disordered! \\

(c) Finally, it was implicitly assumed in the previous section that         
one full electron charge binds to every domain wall unit cell, corresponding with the
filled stripes of the nickelates. However, starting from the more general notion of the sublattice
parity order there is no need to limit oneself to this special case. As a generalization, one might
also want to attach some fraction of the electron charge to the stripe unit-length. This might be half an
electron corresponding with the half-filled stripes of the underdoped cuprates or even an
irrational fraction so that stripes would be undoubtedly internally charge compressible metals -- the metallic
stripes of Kivelson, Emery {\em et. al.} (Kivelson and Emery, 1996; Emery, Kivelson and Zachar, 1997; 
Kivelson, Fradkin and Emery, 1998; Emery, Kivelson and Tranquada, 1999; Carlson {\em et al.}, 2000;
see also, e.g., Castro-Neto and Guinea, 1998; Zaanen, Osman and van Saarloos, 1998;
Voita, Zhang and Sachdev, 2000; Fleck {\em et al.}, 2000; Bosch, van Saarloos and Zaanen, 2001). 
This adds yet another dimension to  the physics and  we
leave a further discussion to these authors. We emphasize, however, that there is a-priori
no conflict between this approach and what is discussed here.  In fact, in the Emery-Kivelson school of thought, the 
local one dimensionality enters as an {\em assumption}. Sublattice parity order offers
a {\em rational} for  this assumption.

\section{Stripes as strings}

\begin{figure}[t]
\begin{center}
\epsfig{figure=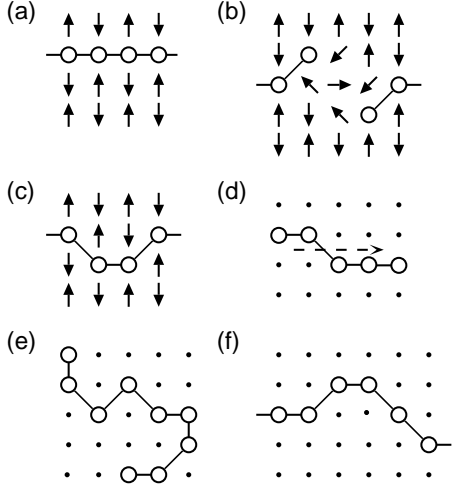,width=6cm}   
\end{center}
\caption{ (a) A charged domain wall separating spin domains 
of opposite AFM order parameter. (b) If domain walls are `broken up' spin-frustrations
emerge and these events are excluded in the mean-field theory of sublattice parity order.
(c) It is still possible to fluctuate the shape of the stripe by constraining its 
trajectory to only nearest- and next-nearest neighbor links. (d) A typical configuration
is the kink and the movements of these kinks are an important source of kinetic energy. (d)
An example of an undirected string configuration. (e) A typical directed string configuration.}
\label{fig_walls}
\end{figure}

Let us take the Ogata-Shiba geometrical squeezing for granted, but now in
2+1D. The requirement that the squeezed lattice is an unfrustrated, bipartite
lattice puts strong constraints on the way the stripes can fluctuate: only 
configurations are allowed where the `holes' form fully connected trajectories,
while every pair of holes is either nearest-neighbor (`horizontal bonds') or
next-nearest-neighbor (`diagonal bonds'), see Fig. \ref{fig_walls}.

 This is
a quite restrictive constraint. However, we notice that it is imposed by
the requirement that sublattice parity is fully ordered. Longer excursions
of holes away from the stripe would cause frustrations in the spin
system on the squeezed lattice, thereby violating the order.
Although in the cuprate reality these are likely
to be quite important, our goal here is to derive the mean-field theory, and for these purposes
it is necessary to neglect these motions. In addition, there
are indications that the long wavelength physics of the single stripe is relatively
insensitive for these `microscopic details'. Hence, starting with
a strong coupling model (with regard to binding of holes to stripes)
we derive a fixed point physics with a finite basin of attraction.

Despite the prescription that holes have to be nearest- or next-nearest-neighbors
it should be immediately obvious that the stripe has still much room to
quantum-fluctuate. What is the nature of the problem?

Because of the connectedness requirement, at every instant of time
the holes have to form 1+1D manifolds, and a single stripe is
therefore a quantized string. As a fortunate circumstance, the physics
of a single string of the type following from the squeezing 
requirement has been already studied in a great detail. Eskes
{\em et al.}, 1996, 1998, introduced precisely this kind of string for  the
purpose of a model study of the stripe fluctuations. Strings are
extended entities which can exist in different collective states.
The theory of  strings, either in the high energy context or the
membranes of statistical physics, is a rich subject  which is far
from completely understood. However, it has long been recognized (Polyakov, 1987)
that strings behave in one regard very  differently from particles.
In the quantum theory of particles, the continuum limit can be
reached by defining the theory on a lattice, taking subsequently
the limit of the lattice constant going to zero. This is in general
untrue for strings and the richness of high energy string theory is
in last instance associated with the true continuum limit. In the
stripe context, the problem starts out with tight binding electrons
moving on the crystal lattice. Hence, whatever else happens, 
the UV is lattice regularized and this  turns out to be a  most important 
condition determining the long wavelength dynamics of the fluctuating
stripe. String theory on a lattice is easy. It coincides with
the statistical physics of crystal surfaces which appears to be
completely understood (van Beijeren and Nolten, 1987).

Eskes {\em et al.} specialized on the particular lattice strings 
which are of direct relevance in the present context: neighboring
`holes' are connected by nearest- or next-nearest links on the
lattice. Subsequently, Morais-Smith {\em et al.}, 1998, ( see also Hasselmann {\em et al.}, 1999;
Dimashko {\em et al.}, 1999) studied a model
with much weaker constraints finding the same fixed points at 
long-wavelength suggesting that the constraints are microscopic
details which are unimportant for the universal long wavelength
physics. Eskes {\em et al.} managed to enumerate all possible
phases of these strings. Although there is a wealth of ordered
or partially ordered phases, corresponding with strings localized
in space due to lattice-pinning, the only delocalizing string
phase corresponds with the Gaussian fixed point, associated with
algebraic long range order. Hence, on the single stripe level
there is already a strong tendency towards localization and
little is needed to  find true long range order in a system of
interacting stripes. To make matters worse, Eskes {\em et al}.
found that at least these particular lattice strings are generically
exhibiting a  rather unusual type of symmetry breaking: these
strings acquire spontaneously a direction in space, even if they
are quantum delocalized. This is like a nematic order: although
translational invariance is restored, rotational invariance is
broken. If the string starts at $-\infty$ at the `left' of the
2D plane, it will always and up at the `right' boundary, and never
at the `upper' or `lower'  boundary.
This `directedness' order originates  in an
order-out-of-disorder mechanism: directedness lowers the kinetic
energy at short distances.
The statement can be made that it always happens because it
works  only better when the fluctuations become more severe.    
    
Let us present  the mathematical
definition of the Eskes strings, as well as a summary of the results
insofar as they pertain to the remainder of this paper.
Consider an configuration of points on the 2D square
lattice, spanning up a 1D trajectory where every pair of points is
connected by either a nearest- (nn) or next-nearest-neighbor (nnn) link
 (Fig. \ref{fig_walls}). 
The coordinate of the $l$-th point is $(\eta^x_l, \eta^y_l)$ and
the string configurations can be projected out from the total space
of these points by,
\begin{equation}
\left| \vec {\eta}(strings) \right> = \prod_{l}{\cal P}({\hat { \vec 
{\eta}}}_{l+1} - 
{\hat { \vec {\eta}}}_{l}) \left| {\vec {\eta}} \right >,
\label{strproj}
\end{equation}
where
\begin{equation}
{\cal P}( {\vec {\eta}}_{l}) = \delta(|  {\vec {\eta}}_{l}| - 1 ) + 
\delta( |{\vec {\eta}}_{l}| -  \sqrt{2} )~,
\label{strproj1}
\end{equation}
ensuring that the neighboring points are not farther apart than
1 or $\sqrt{2}$ lattice constants. The potential energy of the
string can be parametrized by,
\begin{equation}
{\cal H}_{{\cal C}l}^{1} = \sum_{l}{\cal K}\delta (| {\hat \eta}_{l+1}^{x} 
- {\hat \eta}_{l}^{x}| - 1)\delta (| {\hat \eta}_{l+1}^{y} - 
{\hat \eta}_{l}^{y}| - 1),
\label{KpotE}
\end{equation}
expressing that a nnn-link has an energy $K$ relative
to a nearest-neighbor one, and the lattice representation of
curvature energy,
\begin{equation}
{\cal H}_{{\cal C}l}^{2} = \sum_{l}\sum_{i,j=0}^{2} {\cal L}_{ij}\delta (|\eta_
{l+2}^{x} - \eta_{l}^{x}| - i)\delta (|\eta_{l+2}^{y} - \eta_{l}^{y}| - j),
\label{LpotE}
\end{equation}
expressing that e.g. two neighboring nn-links pointing in the
same direction have a different energy than a nnn-link following
a nn-link, or for instance two nn-links pointing in orthogonal
directions. In principle one could also include longer range
link-link interactions but this will not change matters 
qualitatively at long wavelength. The string kinetic energy
is,
\begin{equation}
{\cal H}^{{\cal Q}u} = {\cal T} \sum_{l}{\cal P}_{Str}^{x}(l){\cal P}_{Str}^{y}(l)
\left( e^{i{\hat {\pi}}_{l}^{x}} + e^{-i{\hat {\pi}}_{l}^{x}} + e^{i{\hat 
{\pi}}_{l}^{y}} + e^{-i{\hat {\pi}}_{l}^{y}}\right ), 
\label{strKE}
\end{equation}
where $\hat{\pi}$ is the canonical, periodic lattice momentum 
associated with the position operator $\vec{\eta}$,
\begin{equation}
\left[ {\hat {\eta}}_{l}^{\alpha},{\hat {\pi}}_{m}^{\beta}\right ] = 
i\delta_{l,m}\delta_{\alpha ,\beta}.
\label{strmon}
\end{equation}
Acting once with $\hat{\pi_l^{x}}$ on a string configuration
will cause a hop of point $l$ over a lattice spacing in the
x-direction, as long as the string constraint is not violated.

This model is non-integrable and one can proceed in different
fashions. In the path-integral formalism, a quantum particle
corresponds with a worldline in a one-higher dimensional space,
and likewise a quantum string becomes a worldsheet, a statistical
physics membrane living in 2+1 dimensional embedding space. 
Lattice strings correspond with special membranes, namely those
which also describe the statistical physics of crystal surfaces.
The  role of the lattice in the quantum problem is taken by
the corrugation of the crystal in the crystal surface problem.

It is easily seen that the general form of the action of the
lattice string defined in the above  is that of a {\em restricted Solid-on-Solid}
(RSOS) surface problem. Here the surface is subdivided in
columns with height $\eta_l$ and these column heights interact
via terms like Eq.'s (\ref{KpotE},\ref{LpotE}) expressing that it costs for instance an
energy $K$ to have neighboring columns to differ in height by
one unit, instead of having a flat configuration. It is also not
hard to find out  that the lattice kinetic energy Eq. (\ref{strKE}) acquires a 
similar RSOS form after spreading it out along the time direction.

A specialty of the lattice string is, 
however, that the $\eta^x$ and $\eta^y$ problems are described
separately by their own RSOS surface and the interplay of the motions
along the $x$ and $y$ directions gives rise to strong interactions between
both RSOS `sectors'  via local constraints. For instance,
keeping both surfaces flat amounts to putting all particles
$l$ on the same lattice site. This problem was studied
numerically, using quantum Monte-Carlo, and it was discovered
that in the parameter regime of interest always  directedness
symmetry breaking occurs. A particularly interesting physical  
picture emerges  in the language of  coupled RSOS surfaces. In order to
optimize the freedom to fluctuate, the best the system can
do is to {\em order} one of the surfaces. In doing so, the
constraints coming from the surface-surface coupling 
disappear completely and  the other surface  can
fluctuate freely.  The entropy gained by  this freely fluctuating 
surface out-weights the entropy associated with having
both surfaces disordered. Take the $x$ surface to  be
the ordered one. The order is such that this surface 
always steps upward, corresponding with the string
being directed along the $x$ direction. Along the
$y$ direction the string can now freely quantum 
meander.

\begin{figure}
\begin{center}
\epsfig{figure=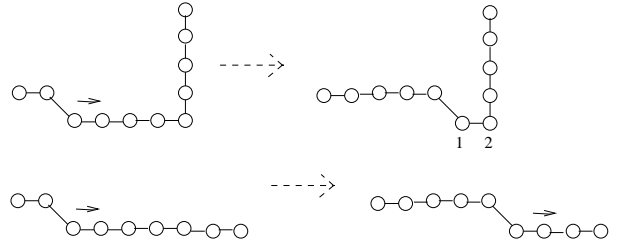,width=8cm}   
\end{center}
\vspace{0.3cm}
\caption{ Illustration of the fact that a bend blocks the 
propagation of kinks along the string. Note that the ` holes' 1 and 2 
adjacent to the bend cannot  move.} 
\label{fig_bends}
\end{figure}

The (directed) string can be viewed as a generalized
quantum sine-Gordon problem, and it is most useful
to consider its physics in terms of its soliton- or
kink degrees of freedom. These are the events as shown
in Fig. \ref{fig_bends} where the string steps sideways. It is
easily checked that under the influence of the kinetic
energy Eq. (\ref{strKE}) these kinks propagate like free
particles and this a dominant source of kinetic energy.
To  gain some intuition in the directedness symmetry
breaking it is instructive to consider what happens when
such a propagating kink approaches an `overhang' in the string,
violating the directedness (Fig. \ref{fig_bends}). It is easily seen that such 
an overhang acts like a hard wall for the soliton. This costs kinetic energy and this 
can be taken as an alternative physical picture for the mechanism 
driving the directedness.

At the same time, this directedness amounts to a great simplification.
The problem is reduced to a single RSOS problem and there is a
great body of knowledge on RSOS-type models. It can be demonstrated
that there are in total 10 distinguishable phases, see table (\ref{tabPhases}).
 Pending
parameters the string can be in various phases dominated by
the potential energy where the stripes are localized in space. E.g.,
the string can be, on average, a straight line, which is pinned by the lattice, 
oriented along the horizontal (phase II), or along a diagonal (phase I)
direction in the lattice. However, also partially ordered phases
are possible (`Haldane', `Slanted' phases) and, last but not least,
there is only one delocalized phase which is Gaussian as stated
earlier.
\setlength{\unitlength}{0.08cm}

\begin{table}[h!]
\hspace{0.5cm} \caption{ A schematic representation of the 10 different
phases of the directed lattice string of Eskes {\em et al.}. Both characteristic
configuration of the strings and that of the equivalent $S=1$ chain are indicated.}  
\begin{center}
\begin{tabular}[H]{ccc}
  Phase  &   \hspace{0.4cm}  String  &  \hspace{0.4cm}  Spin 1   \\
\hline
    I     & 
 \begin{picture}(10,10)
 \multiput(0,0)(2,2){4}{\circle*{0.9}}
\end{picture}
                              &  \hspace{0.4cm}  ++++++++   \\
       II    & 
\begin{picture}(18,4)
 \multiput(0,0)(2,0){9}{\circle*{0.9}}
\end{picture}
                              &  \hspace{0.4cm}  0 0 0 0 0 0 0 0    \\
       III   & 
\begin{picture}(18,6)
 \multiput(0,0)(4,0){5}{\circle*{0.9}}
 \multiput(2,2)(4,0){4}{\circle*{0.9}}
\end{picture}
                              &  \hspace{0.4cm}  +$-$+$-$+$-$+$-$     \\
       IV    &  
\begin{picture}(20,8)
 \put(0,0){\circle*{0.9}}
 \put(2,2){\circle*{0.9}}
 \put(4,2){\circle*{0.9}}
 \put(6,0){\circle*{0.9}}
 \put(8,2){\circle*{0.9}}
 \put(10,2){\circle*{0.9}}
 \put(12,4){\circle*{0.9}}
 \put(14,4){\circle*{0.9}}
 \put(16,2){\circle*{0.9}}
 \put(18,4){\circle*{0.9}}
\end{picture}
                              & \hspace{0.4cm}  +0$-$+0+0$-$+   \\
       V      & 
\begin{picture}(16,6)
 \put(0,2){\circle*{0.9}}
 \put(2,0){\circle*{0.9}}
 \put(4,2){\circle*{0.9}}
 \put(6,2){\circle*{0.9}}
 \put(8,0){\circle*{0.9}}
 \put(10,0){\circle*{0.9}}
 \put(12,0){\circle*{0.9}}
 \put(14,2){\circle*{0.9}}
 \put(16,0){\circle*{0.9}}
\end{picture}
                              & \hspace{0.4cm}  $-$+0$-$0 0+$-$     \\
       VI      &  
\begin{picture}(16,8)
 \put(0,0){\circle*{0.9}}
 \put(2,2){\circle*{0.9}}
 \put(4,0){\circle*{0.9}}
 \put(6,2){\circle*{0.9}}
 \put(8,4){\circle*{0.9}}
 \put(10,2){\circle*{0.9}}
 \put(12,4){\circle*{0.9}}
 \put(14,2){\circle*{0.9}}
 \put(16,0){\circle*{0.9}}
\end{picture}
                              & \hspace{0.4cm}  +$-$++$-$+$-$$-$     \\
       VII     &  
\begin{picture}(10,10)
 \put(0,0){\circle*{0.9}}
 \put(2,2){\circle*{0.9}}
 \put(4,2){\circle*{0.9}}
 \put(6,4){\circle*{0.9}}
 \put(8,6){\circle*{0.9}}
 \put(10,6){\circle*{0.9}}
\end{picture}
                              & \hspace{0.4cm} 0+0++0+0 0     \\
       VIII     &  
\begin{picture}(14,10)
 \put(0,0){\circle*{0.9}}
 \put(2,0){\circle*{0.9}}
 \put(4,2){\circle*{0.9}}
 \put(6,2){\circle*{0.9}}
 \put(8,4){\circle*{0.9}}
 \put(10,4){\circle*{0.9}}
 \put(12,6){\circle*{0.9}}
 \put(14,6){\circle*{0.9}}
\end{picture}
                              & \hspace{0.4cm}  0+0+0+0+0     \\
       IX       & 
\begin{picture}(18,8)
 \put(0,0){\circle*{0.9}}
 \put(2,0){\circle*{0.9}}
 \put(4,2){\circle*{0.9}}
 \put(6,2){\circle*{0.9}}
 \put(8,4){\circle*{0.9}}
 \put(10,4){\circle*{0.9}}
 \put(12,2){\circle*{0.9}}
 \put(14,2){\circle*{0.9}}
 \put(16,4){\circle*{0.9}}
 \put(18,4){\circle*{0.9}}
\end{picture}
                              & \hspace{0.4cm}  0+0+0$-$0+0     \\
       X      &   
\begin{picture}(14,6)
 \put(0,0){\circle*{0.9}}
 \put(2,0){\circle*{0.9}}
 \put(4,2){\circle*{0.9}}
 \put(6,2){\circle*{0.9}}
 \put(8,0){\circle*{0.9}}
 \put(10,0){\circle*{0.9}}
 \put(12,2){\circle*{0.9}}
 \put(14,2){\circle*{0.9}}
 \put(16,0){\circle*{0.9}}
 \put(18,0){\circle*{0.9}}
\end{picture}
 & \hspace{0.4cm}  0+0$-$0+0$-$0     
\end{tabular}
\end{center}
\protect\label{tabPhases}
\end{table}

To get more insight in this phase diagram, it is instructive to consider yet another
representation of the problem:  the directed string corresponds
with a  $S=1$ Heisenberg spin chain with added Ising and single site anisotropies.
 This is easily seen in terms
of a representation where the {\em links} are the dynamical degrees
of freedom. Single out a particular `guiding point' $\eta_0$ on
the directed string and it is immediately clear that the string
dynamics can be completely parametrized in terms of its center of
mass $\eta_0$ and the relative coordinates corresponding with the
set of link variables taking the values $1, 0, -1$ corresponding
with (1,1), (1,0) and (1,-1) bonds, respectively, for a string
directed in the x direction. For an infinitely long string the center
of mass coordinate becomes non-dynamical, and the problem is completely parametrized 
in terms of the   
possible states on the links. These can be as well viewed
as the three $M_S=1,0,-1$ states of a $S=1$ quantum spin. For instance,
the string kinetic energy is equivalent to the XY term in the $S=1$
spin representation, $\sim S^+_l S^-_{l+1} + h.c.$ because 
$S^+ = \sqrt{2} ( | 1 \rangle \langle 0 | + | 0  \rangle \langle -1 |)$
in the basis of eigenstates of microscopic spin. Hence, acting once
with this term changes two horizontal links into the two diagonal links
corresponding with the sideward motion of the hole in the middle.

The famous Haldane phase of the Heisenberg $S=1$ spin chain has a
particular simple interpretation in the string language 
where it corresponds with a form of partial order (den Nijs and Rommelse, 1989). 
In this phase (V in table I), kinks have proliferated in the ground state and in this regard the state
is quantum disordered. However, there is still a form of hidden
order in the sense that at average every kink which is moving the
string upward is followed by a kink which is moving the string
downward. Hence, the string as a whole is still localized in
space although it is now localized in the middle of two neighboring
rows of the lattice (like a `bond-ordered' stripe). This type of
order is hidden from the spin-correlators and to make it visible
in the spin chain one needs a non-local correlator. Eskes {\em 
et al}. discovered also a second type of partially ordered string:
the `slanted string' (VII). This is like the Haldane phase except that
the kink `flavors' are now ferromagnetically ordered such that the
string orders along an arbitrary direction in the lattice. It
was recently suggested that such a phenomenon might be relevant
in the cuprate context (Bosch, van Saarloos and Zaanen, 2001).

Most importantly, it is well established that $S=1$ quantum spin chains
have only a single massless fixed point whose basin of attraction
includes the $XY$ point (phase IV), where the string only has kinetic energy. 
This is a Gaussian fixed point and this
is the only phase where the string is delocalized in space.
This is  an exceedingly simple fixed point: at large distances,
the motions of the string can  be completely parametrized
in terms of the non-interacting transversal phonon-modes of the string. The 
position of  points on the string can be written as $\eta (l) =
\eta_0 (l) + \delta \eta (l)$, where $\eta_0$ corresponds with the
position of a flat string while $\delta \eta$ corresponds with the
transversal displacement. Following the standard lore of Gaussian
theory it follows that the displacement correlator diverges 
logarithmically $\langle (\delta \eta (l) - \delta \eta (0) )^2 \rangle
\sim ln (l )$ such that the string density correlator decays
algebraically $ \langle \rho (l) \rho (0) \rangle \sim 1 / l^K$.

The conclusion is, that a single string is either ordered or algebraically ordered.
If one insists that stripes are fully connected entities, this might
well be a very general conclusion. A major limitation of the above
work is that it assumes the stripe to be internally
incompressible and it is a-priori unclear what happens when the
stripes are internally metallic. A specific string model has been 
constructed for a metallic stripe by ourselves (Zaanen, Osman and van Saarloos, 1998).
 Although this also shows algebraic order at best, specific
assumptions are made in its construction which render it to be less
general and it might not be representative for the general case.

However, on  general terms it is not easy to see how one 
can avoid the algebraic string order. A sufficient condition for
algebraic order is the non-vanishing of the line tension. How to
get rid of a term in the action which is proportional to the
world sheet area? There is actually one possibility which deserves
a further exploration (Mukhin, van Saarloos and Zaanen, 2001): 
let the stripes be in equilibrium with a
quantum gas of charges, while these charges can freely enter and
leave the stripes. Under these conditions stripes are described by
a qualitatively different  type of strings: the so-called extrinsic
curvature strings. These are well known in the statistical physics
context. Consider for instance biological membranes.
 These membranes  are  immersed in a `gas' of constituents (lipids) which is in
equilibrium with the membranes. If one pulls the membrane
it  simply absorbs lipids without paying a free energy penalty and
therefore the tension (proportional to the membrane area) vanishes. 
Instead, the next order invariant
takes over, corresponding with extrinsic curvature. After 
linearization, the action $S \sim (\delta^2_{\mu} \eta )^2$, instead
of the finite tension case $S \sim (\delta_{\mu} \eta )^2$, and from
power counting one infers directly that such an extrinsic curvature
membrane fluctuates in the same way as a line with tension
(like a worldline). The analogy with the stripe immersed in a
bath of free charges should be immediately clear. There is however
a caveat. The bare particles are non-relativistic and their 
action contains a mass term $\sim ( \partial_{\tau} \eta )^2$.
Since the stripes are supposed to be made out of these particles
it is hard to see how one can get rid of this tension in the
time direction, even when tension vanishes on the time-slice.

\section{Order out of disorder in the system of stripes.} 

The main conclusion of the previous section is that 
a single quantum stripe, as defined through the sublattice parity order,
is at best a very mildly fluctuating object. Considering a system of
these Gaussian strings, at the moment one adds any interaction it has
to be that long range order sets in.  Algebraic
order (of the single stripe) changes in  true order in the presence of any perturbation,
regardless its strength. Only 
recently exceptions have been identified (the quantum smectic, or
gliding phase, Kivelson, Fradkin and Emery, 1998; Emery {\em et al.}, 2000).
 However, these are only realized under specific
circumstances which are not found in the present context.

Hence, any direct interaction between the stripes suffices to 
cause translational symmetry breaking in the system of stripes.
There is no doubt that the stripes are interacting. They are
charged and therefore they should exert Coulomb forces. In
addition there are the Casimir-type forces in the spin system (Pryadko, Kivelson and Hone, 1998),
as well as the elastic forces mediated by the lattice.

Although  mostly of  academic interest, hard-core
interactions (or non-intersection conditions) are special (Zaanen, 2000). These
interactions are highly singular and a priori one cannot be sure
that the hard-core interaction will play the same role as finite range
interactions. Although there is no real good reason, it is appealing
to assume that the stripe-stripe interaction contains a hard-core 
piece. One might want to be interested specifically in the question
to what extent can stripes be a one dimensional sub-reality in two
dimensional space. For this purpose alone one would like to keep
stripes from intersecting. In addition, if one just wants to generalize
Ogata-Shiba to one higher dimension, one also better keep their hard
cores attached to the charges. 

As discussed in section II, the charge sector of the Luttinger
liquid of a strongly coupled Hubbard model is described by a
hard-core bose gas. A most literal generalization of this
Luttinger liquid to 2+1D can be obtained by just a {\em
decompactification}, in the same sense as used in
fundamental string theory. 
In path-integral language, the hard-core
bose gas corresponds with meandering elastic worldlines, directed
along the time dimension, which cannot intersect (hard-core
condition). At distances large compared to the lattice constant $a$
one cannot see the difference between this system and a system
of strings characterized by  one more space dimension which is 
curled up in a circle with compactification radius $R \simeq a$,
with the string wrapped around this extra dimension. 
Decompactification means that the compactification radius
$R \rightarrow \infty$. What happens? The tiny string cylinders
spread out in 2D worldsheets, corresponding with elastic membranes, spanning the
extra space dimension. The hard-core condition means that these
worldsheets cannot intersect. This entity was called the {\em directed}
string gas in 2+1D. The emphasis should be on directed because 
this decompactification construction gives rise to a constraint
which is a-priori not completely general. On the time-slice the
strings are {\em directed} along the extra dimension. A 
string starting at $-\infty$ in this direction always ends up
at $+\infty$ in the same direction.

It is a fundamental requirement of non-relativistic quantum-mechanics
that worldlines/worldsheets are directed along the time direction.
However, no general constraint of this kind acts in space directions,
and in principle `overhangs' or `dislocations' (Fig. \ref{dislocation}), where a string
for example starts out at $-\infty$ to end at $-\infty$ (for
open boundaries), are in principle possible. 

A first objection could be that a single string can be subjected
to directedness symmetry breaking, the surprise in the previous
section. If the constituents are already directed, the system
will be definitely directed. However, although it is demonstrated
that lattice strings of the previous section can acquire spontaneously a direction,
there is no theorem available stating that lattice strings are
{\em always} directed. Hence, one cannot claim that lattice
string gasses are universally directed.

However, there is an elegant argument available demonstrating
that directedness is an unavoidable consequence of the dynamics
in the {\em system} of hard-core elastic strings. This goes hand
in hand with the demonstration that the directed string gas
has to solidify (to break translation symmetry) {\em always}.
Exceptions are not possible. Hence, together with the physics
discussed in the previous section, the conclusion is that if the
Ogata-Shiba geometric squeeze prescription applies literally,
long range order is unavoidable at zero temperature in 
2+1 dimensions! 

Let us discuss the string-gas in more detail. The theoretical
problem is that due to the absence of a second quantization 
formalism the canonical methods of quantum mechanics are of no
use for string problems. Hence, all what remains is the path
integral formalism and in this formalism the string-gas problem
corresponds with the statistical physics problem of elastic
membranes embedded in 3D space subjected to a non-intersection
condition, with the added constraint that the membranes are
directed along one (imaginary time) direction.

Let us step back, to reconsider the (seemingly) easier `compactified'
version corresponding with directed, non-intersecting elastic lines 
in 2D. This is equivalent to the 1+1D hard-core bose gas and it
is well known that this is in turn equivalent to the problem of
non-interacting spinless fermions in 1+1D. This is of course
a trivial problem and the freshman can calculate the density-density
correlator of the fermion gas to find,
\begin{equation}
\langle n(r)n(0) \rangle = -\frac{2}{(\pi r)^2} + \frac{2 \cos(2k_F r)}{(\pi r)^2}
\label{fermgas}
\end{equation}
and the textbook will stress that these are the famous 
Friedel-oscillations, characteristic for any fermi-gas
in any dimension.

However, much later one learns that the spinless-fermion gas
is just a Luttinger  liquid characterized by a charge stiffness
$K_{\rho} = 2$. In turn, since the observations by Haldane (Haldane, 1981)
and others it is clear that Eq. (\ref{fermgas}) has to do with algebraic
long range order. Hence,  the bosons order in a
1+1D crystal. This crystal is carrying phonons and the admixture
of these phonons in the ground state change the true long range
order in the algebraic order signaled by Eq. (\ref{fermgas}). This
appears as a paradox: the Fermi-gas is mere kinetic energy and how
can this gas ever be a crystal? The resolution is that 
Fermi-statistics codes for a hard-core condition in the Bose
language, and the hard-cores cause microscopic kinetic energy
to become potential energy at large distances, driving the order. 
An interesting order-out-of-disorder mechanism is hidden 
behind the simple non-interacting fermions!

This mechanism is well known in the statistical physics,
addressing the problem of classical incommensurate
fluids (domain wall fluids) in 2D (Pokrovsky and Talapov, 1979;
Coppersmith {\em et al.}, 1982). 
The argument goes back to
work by Helfrich, 1978,  actually on extrinsic curvature membranes in
2+1D, and was apparently reinvented in the community working
on 2D incommensurate fluids. In the 1+1D context one can either
use an intuitive argument or a more rigorous self-consistent 
phonon method invented by Helfrich. In 1+1D one arrives at
the same answer (at least qualitatively) but this is different
for elastic strings in 2+1D, where  the
intuitive argument is flawed. Nevertheless, the intuitive 
argument is instructive because it sheds light on the basic
physics at work.

This arguments is as follows for the 1+1D case. 
The hard-core bose gas at zero temperature
corresponds with the statistical physics problem of a gas of non-intersecting
elastic lines embedded in 2D space-time, which
are directed along the time direction. The space-like 
displacement of the $i$-th worldline is parametrized in terms of a field
$\phi_i (\tau)$ ($\tau$ is imaginary time) and the partition function
is ($M$ is the mass of the particle), 
\begin{eqnarray}
Z & = & \Pi_{i=1}^N \Pi_{\tau} \int d \phi_i (\tau) 
e^{-{ S \over {\hbar}} }, \nonumber \\
S & = & \int d\tau \sum_i {{M} \over 2} ( \partial_{\tau}
\phi_i)^2, 
\label{hcZ}
\end{eqnarray}
supplemented by the avoidance condition,
\begin{equation}
\phi_1 < \phi_2 < ...  < \phi_N.
\label{avoid}
\end{equation} 
The hard-core condition Eq.(\ref{avoid}) renders this to be a highly
non-trivial problem.

At short distances the worldlines can freely meander. However, after
some characteristic time-like distance, the
worldlines will collide. In the statistical physics analogy, every
collision costs an entropy $\sim k_B$ because the lines cannot intersect.
Hence, these collisions raise the free energy of the system and
this characteristic free energy cost $\Delta F_{coll} \sim
k_B T n_{coll.}$. The density of collisions $n_{coll.}$ is easily
calculated for  the elastic  worldlines. It follows from equipartitioning
that  the mean-square transversal
displacement as function of (time-like) arclength increases
like $\langle \left[ \phi  (\tau) - \phi (0) \right]^2
 \rangle = ( \hbar / M) \tau$.
The characteristic time $\tau_c$ it takes for one collision to occur is
obtained by imposing that this quantity becomes of order $d^2$ where
$d$ is the average worldline separation, while the particle
density $ n \sim 1/d$.
A characteristic collision energy scale is obtained $E_F \sim 
\hbar / \tau_c \sim (\hbar^2/ M) n^2$.
$E_F$ is of course the Fermi-energy: it is the scale separating a
regime where worldlines are effectively isolated ($E > E_F$, free
particles) from the one dominated by the collisions ($E < E_F$, Luttinger
liquid). 

At the same time, the entropy/kinetic energy cost gives rise to
an effective {\em repulsion} between the world-lines, and this
repulsion is in turn responsible for the ordering tendency. At large
distances the precise origin of the repulsion does not matter and
one can simply assume that the entropic repulsion is like a
harmonic spring and the  spring constant can be estimated 
by taking the ratio of the characteristic energy  ($E_F$) and the
characteristic distance $d$. In this way one finds a `induced
modulus' $B$ associated with the compression of the hard-core 
1+1D quantum  gas,
\begin{eqnarray}
B_0 & \sim & E_F / d \nonumber \\
    & \sim & { {\hbar^2} \over {M d^3} }.
\label{est}
\end{eqnarray}
Asserting that at long wavelength the gas is described by the elasticity
theory of a 1D quantum crystal with spatial modulus $B_0$ and mass density $\rho \sim M / d$,
\begin{equation}
S_{eff}  = { 1 \over 2} \int d\tau \int dx 
\left[ { \rho} (\partial_{\tau} \psi)^2 + B_0 (\partial_{x} \psi)^2 \right],
\label{BSeff}
\end{equation}
one recovers the spinless-fermion results, modulo prefactors of order unity.

The more rigorous argument by Helfrich, 1978, starts out by assuming that the
Bose-gas is described by the long wavelength action Eq. (\ref{BSeff}). 
In the absence of the hard-core interaction $B_0$ would be zero by
definition and the free energy increases for a finite $B_0$
because the fluctuations are suppressed. Define a `free-energy
of membrane joining' as $\Delta F = F (B_0) - F (B_0 = 0)$. At the
same time, by general principle it has to be that the true modulus
in the space direction $B$ should satisfy ($V$ is the volume),
\begin{equation}
B = d^2 { { \partial^2 (\Delta F ( B_0) / V)} \over {\partial d^2} }. 
\label{diffeq}
\end{equation} 
In case of the steric interactions, the only source of long wavelength
rigidity is the fluctuation contribution $\Delta F$. Therefore 
$B = B_0$ and $B$ can be determined self-consistently from the
differential equation Eq. (\ref{diffeq}). This method is not
exact, because mode couplings are neglected. However, these
mode couplings are important at short distances and they are
therefore not expected to change the outcomes qualitatively.
The ultraviolet only enters the answers through the short
distance cut-off in the integrals, $x_{min} = \eta d$ and
it appears that all the effects of these interactions can be
absorbed in the fudge factor $\eta$. Evaluating matters
for the hard-core Bose gas, it turns out that it reproduces exactly
the spinless fermion results if $\eta = \sqrt{6}$ (Zaanen, 2000). 

The conclusion is that the algebraic translational order hidden
in the hard-core Bose gas/spinless fermion problem in 1+1D can
be understood as an order-out-of-disorder phenomenon in the
equivalent statistical physics problem, which can be handled 
rather accurately, using a simple statistical physics method. 
The advantage is that the Helfrich method applies equally well
to the string gas problem in 2+1D. In fact, it works even better!

Let us first consider the directed string gas. The bare action
of this string gas 
in Euclidean space-time describes  a sequentially ordered stack 
of elastic membranes. Orienting the worldsheets in the $y, \tau$ planes,
the action becomes in terms of the dispacement fields $\phi_i(y,\tau)$
describing the motion of the strings in the $x$ direction,  
\begin{eqnarray}
Z & = & \Pi_{i=1}^N \Pi_{y,\tau} \int d \phi_i (y, \tau) 
e^{-\frac{S}{\hbar} }, \nonumber \\
S & = & \int d\tau dy \sum_i \left[ {{\rho_c} \over 2} ( \partial_{\tau}
\phi_i)^2 +  {{\Sigma_c} \over 2} ( \partial_y \phi_i )^2 \right], 
\label{strZ}
\end{eqnarray}
again supplemented by the avoidance condition Eq. (\ref{avoid}).
In Eq. (\ref{strZ}), $\rho_c$ is the mass density and $\Sigma_c$ the string 
tension, such that $c = \sqrt{ \Sigma_c / \rho_c}$ is the velocity. 

Let us now  consider the intuitive collision-argument for this string gas. 
The mean-square transversal displacement now depends 
logarithmically on the worldsheet area $A$: 
$\langle (\Delta \phi (A))^2 \rangle = \hbar /(\rho c) \ln (A)$.
Demanding this to be equal to $d^2$, the degeneracy scale 
follows immediately. The characteristic worldsheet area $A_c$
for which on average one collision occurs is given by 
$\hbar/(\rho c) \ln (A_c) \simeq d^2$
where $A_c = c^2 \tau^2_c / a^2$ in terms of the collision time $\tau_c$.
It follows that $\tau_c \simeq (a/c) e^{1 /2\mu}$ and the `Fermi energy'
of the string gas is of order
 $E_F^{str} = \hbar / \tau_c \simeq (\hbar c / a)
\exp { (-1 / 2\mu)}$  where
$\mu$ is the coupling constant (`dimensionless $\hbar$') of the string gas
(Zaanen, Horbach and van Saarloos, 1996),
\begin{equation}
\mu = { {\hbar} \over { \rho c d^2} }.
\label{mu}
\end{equation} 
For a continuum description to make sense, $\mu < 1$ and this suggests that
the Fermi energy  is exponentially small. However, it is finite
and this is all what matters as we will see.

One could be tempted to estimate the induced modulus by asserting $B \sim E_F^{str}$.
However, contrary to the Bose-gas the above intuitive argument is qualitatively flawed and the
reason is that the kinetic repulsions are no longer driven by the  physics at the collision
length scale. A single string is itself a 1+1D elastic entity, characterized by long
wavelength fluctuations which are dangerous in the sense that these are responsible
for changing true long range order in algebraic long range order. Helfrich's self-consistent
phonon method shows that these {\em long-wavelength} single string fluctuations are also the ones
responsible for the induced modulus (Zaanen, 2000). Carrying out the integrations one finds for the 
free-energy of membrane joining,
\begin{equation}
{ {\Delta F} \over V} = { { \pi \hbar c} \over {24 \eta^3 \Sigma_c} }
( { B \over {d^2} }) ( { 5 \over 3} +  \ln \left[ 
{ { \eta^2 \Sigma_c} \over {a^2} } { d \over B } \right] ) + 
O (\lambda^4).
\label{delFsa}
\end{equation}
expanding matters in the small parameter $\lambda = ( \sqrt{  B }  a )/ (\sqrt{\Sigma} \eta d )$.
Since $B$ is tending to zero, the logarithm is dominating and this  term originates
in the small momentum cut-off (long wavelength limit) in the integration of the on-string fluctuations. 

The differential equation obtained by inserting Eq. (\ref{delFsa}) in the self-consistency condition
Eq. (\ref{diffeq}) can be solved and this yields,
\begin{equation}
B = A d^2 e^{ - \eta ({ 54 \over {\pi}})^{1/3} { 1 \over {\mu^{1/3} } } },
\label{Bfinal}
\end{equation}
where $A$ is an integration constant while $\mu$ is the coupling constant defined in Eq. (\ref{mu}).
Hence, instead of the exponential of the `naive' argument,  a stretched exponential is found and this
difference is entirely due to the logarithm in Eq. (\ref{delFsa}), finding its origin in the long
wavelength on-string fluctuations. Hence, it is in this sense that
the solidification of the string gas is driven by the longest wavelength string fluctuations. 

Although the induced modulus is larger than naively expected, from a more practical viewpoint it
is still quite small and it tends to be overwhelmed by the effects of finite range interactions.
This reflects of course the fact that strings fluctuate much less than particles.
However, we set out to demonstrate that long range order cannot be avoided in the string gas and for
this purpose all what matters is that the modulus $B$ is finite at zero temperature. {\em This
is a sufficient condition to exclude a zero-temperature proliferation of dislocations.} In the
absence of the dislocations (Fig. \ref{dislocation}) the string gas is spontaneously directed and the directed gas 
solidifies always, as we showed in the previous paragraphs.

The argument that dislocations cannot proliferate at zero temperature is quite nontrivial (Pokrovsky and Talapov, 1979;
Coppersmith {\em et al.} 1982).
The string-gas theory Eq. (\ref{strZ}) is generalized to finite temperature by compactifying
the imaginary time axis with radius $R_{\tau} = \hbar / k_B T$. The non-proliferation theorem
follows directly from the well-known result that a  Kosterlitz-Thouless transition (dislocation
unbinding driven by thermal fluctuations) happens in this classical string gas  at a {\em finite temperature 
as long as the zero-temperature modulus is finite.} Hence, dislocations are already bound at a finite
temperature and they remain to be bound at zero temperature.

A detailed analysis of the finite temperature case will be presented elsewhere (Mukhin, van Saarloos and Zaanen, 2001). 
The bottomline is that at finite temperatures one can
simply use the high temperature limit of Eq. (\ref{strZ}) (without the time direction), adding however
the induced zero-temperature modulus $\sim B ( \phi_i - \phi_{i+1})^2$. This is nothing else than again
the hard-core bose gas but now in its classical interpretation of thermally fluctuating elastic lines.
The qualitative  difference with the quantum case is that there is no longer a directedness constraint
on the lines and in this classical gas dislocations can occur. If $B = 0$ the remarkable
result is that {\em at any finite temperature dislocations are proliferated}, while at the same
time {\em for any finite $B$ the Kosterlitz-Thouless temperature occurs at a finite temperature
$T_{KT} \sim B$}.

This has been discussed elsewhere at great length (Coppersmith {\em et al.}, 1982)
and let us just repeat the essence of the argument.
The dislocations interact with long range, logarithmic forces which are set by the elastic moduli of
the medium and therefore the energy associated with free dislocations is logarithmic in the system size.
At the same time, the entropy associated with free dislocations is also logarithmic and balancing these
two yields the Kosterlitz-Thouless criterion for the stability of the algebraic order,
\begin{equation}
{ { a d \sqrt{ B_T \Sigma_c } } \over { 2 \pi T } } >  1
\label{KT}
\end{equation}
Using the transfer-matrix the induced modulus $B_T$ of the classical problem can be calculated exactly. Modulo
prefactors this is Eq. (\ref{est}) expressed in classical units. One finds $\sqrt{ B_T \Sigma_c} \sim T$ which means that
either the KT criterion is never satisfied (meaning that dislocations are always bound) or that the
KT criterion is always satisfied so that dislocations are proliferated at all temperatures. It turns
out that the prefactors conspire in such a way that for two flavors of domains (our case) the second 
possibility is realized. This means that at any finite temperature dislocations always proliferate  
but they do so in the most marginal way.
The entropic interactions driven by the finiteness of temperature are on the verge of beating the entropy of the dislocations but
the former just loose. Any interaction other than this entropic interaction
(including the quantum `entropic' interaction) can tip the balance (Mukhin, van Saarloos and Zaanen, 2001). Hence, adding a
finite zero temperature $B$ causes the Kosterlitz-Thouless temperature to happen at a finite temperature.

The conclusion is, remarkably, that nothing can keep the string gas away from solidifying at zero 
temperature (Zaanen, 2000).

\section{The quantum magnetism of static stripes.}

The magnetism of the stripe phase is relatively easy to study experimentally, and for this
reason it is a relatively well developed subject. To put the remainder of this section in
an appropriate perspective let us therefore start out with a sketch of the present empirical
picture.  

It is a rather significant empirical fact that despite a high  hole density  well-developed
antiferromagnetic order can be realized in doped cuprates (Tranquada {\em et al.}, 1995;
Klauss {\em et al.}, 2000). It is well understood that 
a single hole is a strongly frustrating influence in the quantum-antiferromagnet. Since the
spin system itself is quite quantum-mechanical ($S=1/2$) these frustrations give rise to
the formation of a droplet of quantum spin liquid surrounding the hole (Dagotto, 1994). If these holes
would stay independent,  antiferromagnetic order would
disappear at a very low doping. The very fact that N\'eel order has been demonstrated 
to persist in some systems to dopings as large as 20 \% (Klauss {\em et al.}, 2000)
 should be taken as the leading
evidence for the hypothesis of Section II. Of course, it is also experimental fact that this
big N\'eel order occurs when the charges organize in the stripes. However, in doing
so the spin system becomes unfrustrated and this should be understood as the
manifestation of the Ogata-Shiba squeezing principle at work in 2+1D. 

However, on closer inspection one finds that the stripe-antiferromagnet is a more
quantum-mechanical entity than the antiferromagnet of the half-filled insulator.
Both NMR measurements (Hunt {\em et al.}, 1999; Curro {\em et al.}, 2000;
Teitelbaum {\em et al.}, 2001) and neutron scattering (Tranquada, Ichikawa and Uchida, 1999)
indicate that the spin-stiffness
is smaller than the one at half-filling. It has been claimed that this should be due to
a dilution effect: the exchange bonds connecting spins on opposite sides of
the stripes ($J'$) would be very small as compared to the exchange interactions
inside the magnetic domains ($J$) which are in turn believed to be of the same 
magnitude as the exchange interactions at half-filling. However, for
several reasons this cannot be quite the case. First, $J'$ sets the scale for the
overall incommensurate behavior and at energies larger than $J'$ incommensurate
spin fluctuations  cannot exist (Zaanen and van Saarloos, 1997). These fluctuations
have been seen up to energies of $\sim 40 meV$ (Aeppli {\em et al.}, 1997; Mook {\em et al.}, 2000)
and this sets a lower bound
to the value of $J'$. More directly, some inelastic neutron scattering
data are available for the spin waves in a static stripe phase and these
demonstrate that although the stiffness is strongly reduced the spin wave
velocity stays large (Tranquada, Ichikawa and Uchida, 1999). 
This behavior is characteristic for the generic
long wavelength physics of a N\'eel state which is on the verge of
undergoing a quantum phase transition into a quantum-disordered state
(Sachdev, 1999, 2000).

The above observations are associated with the 214 system. Recently,
Mook {\em et al}, 2001, reported evidence for static stripes in the strongly
underdoped 123 cuprates. However, they also claimed that although charge
order is established, the spin system is apparently quantum disordered.
The incommensurate spin fluctuations are seen only above a small but
finite ($\sim 3$ meV) energy. This is not surprising.
It is well understood that the bilayer couplings as they occur in 123
are a factor promoting quantum spin fluctuations (Millis and Monien, 1993; 
van Duin and Zaanen, 1997). Since the spin system
in the single layer 214 cuprates is already on the verge of quantum-melting,
these bilayer couplings could easily tip the balance.

NMR measurements have shown that the actual asymptotic spin-ordering 
process is highly anomalous (Hunt {\em et al.}, 1999; Curro {\em et al.}, 2000;
Teitelbaum {\em et al.}, 2001). It appears that slow spin fluctuations
(MHz scale) show up at the temperature where the scattering experiments
indicate a freezing behavior ($\sim 70$ K), to continue down to the
lowest measured temperatures ($400$ mK). These fluctuations are at
present not at all understood. However, although the case is definitely
not closed, it appears that the spin dynamics on a larger energy scale 
fits quite well the expectations of the generic field theory describing
the long wavelength dynamics of a collinear quantum-antiferromagnet
close to its quantum phase transition. All what matters is the symmetry
of the order parameter ($O(3)$) and the dimensionality of space-time:
this generic theory is the $O(3)$ quantum non-linear sigma model in
2+1 D (QNLS).

Several excellent treatises are available, both on the introductory (Sachdev, 2000) and
the advanced level (Chakravarty, Nelson and Halperin, 1989; Sachdev, 1999), 
on the physics near quantum phase transitions. Let us
therefore limit ourselves to the bare essence. It is well understood 
that the non-frustrated Heisenberg quantum-antiferromagnet defined on
a bipartite lattice does not suffer from Marshall sign problems. Stronger,
the long wavelength dynamics in the semi-classical regime is free of
Berry-phases and it can therefore be described with the simple  QNLS (Fradkin, 1991),
\begin{eqnarray}
Z & = & \int {\cal D} \vec{n} \delta(|\vec{n}| -1 ) e^{-S} \nonumber \\
S & = & {1 \over g_0} \int d^2 x \int_0^{\beta} d\tau  (  (\partial_{\tau} \vec{n})^2 + (\nabla \vec{n})^2 )
\label{QNLS}
\end{eqnarray}
in scaled variables, such that the spin wave velocity is one. $\vec{n}$ is
a three component vector of fixed length and Eq. (\ref{QNLS}) is nothing
else than the theory of a {\em classical} Heisenberg spin system embedded in
2+1 dimensional Euclidean space time (Chakravarty, Nelson and Halperin, 1989). 
At zero-temperature ($\beta \rightarrow
\infty$) this becomes precisely equivalent to the classical Heisenberg problem
in 3D. Hence, for small bare coupling $g_0$ (low temperature in the classical
problem) N\'eel order is established. At a critical value $g^0_c$ a second order
phase transition occurs to a strong coupling, quantum disordered state. In the
classical problem the spin correlators decay exponentially in the disordered
state and this means that the real time dynamics of the quantum problem is characterized
by a dynamical mass gap in the mode spectrum. Right at the critical point the
dynamics is scale invariant while in the proximity of this point the same is true 
up to a length/time scale (Josephson correlation length) where the system finds out that
it either gets attracted to the N\'eel fixed point or to the disordered state. At higher
energies and larger momenta the system still behaves as if it is at its critical point,
until a crossover is reached below which one sees the dynamics associated with the
N\'eel state (zero-modes) or the disordered state (massive triplet excitations). 
The behavior at finite temperature is especially interesting. Finite temperature
means compactification of the imaginary time axis with compactification radius $\hbar / k_B T$. In
the critical regime this breaks scale invariance meaning that the dynamics is
characterized by a characteristic time and length $\tau_c \sim l_c \sim \hbar / k_B T$.

The question arises what this means for the real time dynamics -- the Wick rotation
is a remarkably counter-intuitive affair. The answer is  Sachdev's achievement (Sachdev, 1999). At
times $\tau < \tau_c$ the quantum field theory itself generates through the analytic
continuation {\em a classical relaxational dynamics} characterized by the `quantum
limit of dissipation': this relaxation time is as short as it can be, namely
$\hbar / k_B T$. At shorter times the zero-temperature critical dynamics is recovered,
characterized by a cusp-like dynamical susceptibility reflecting the criticality in
space-time. Away from the critical point, it is the same at high temperature, high
frequency and large momenta, but at temperatures low compared to the Josephson scale 
the mode-excitations associated with the stable fixed points dominate the low-frequency
end of the spectra.

This theoretical picture seems to explain the most salient features of the spin dynamics
in the cuprates, at least if one identifies this spin dynamics with the incommensurate
spin phenomena which are most naturally interpreted as being related to the stripe
N\'eel state. The spin dynamics of the static stripes would then be interpreted as reflecting the
classical sector, that of the fully developed superconductors at low temperatures with that
of the  disordered massive state  (characterized by  a spin gap in the incommensurate spectrum),
while the normal state well above the ordering temperatures would be related to the critical
regime. A crucial assumption is of course that some form of spin-charge separation takes
place and this is in fact clearly excluded by the experiments: the spin gap appears
at the superconducting transition. Hence, this interpretation is at best only part of
the explanation. 

However, in the context of static stripes the situation is more
clear. The interpretation in terms of the stripe-antiferromagnet being on the verge
of undergoing the quantum-phase transition rests on data obtained below the
charge ordering temperature, and in this context a spin-only interpretation is
more reasonable, while it adds credibility to the notion that QNLS has something
to do with high Tc.

The next question is, what is the source of the strong quantum fluctuations in the
stripe antiferromagnet? This can have several reasons. A first obvious possibility
is, in the language of this paper, `local violations of the Ogata-Shiba squeeze' --
by longer excursions away from the stripes holes can cause local violations of the
bipartiteness of the squeezed lattice. These correspond with local spin-frustrations 
and  they should therefore enhance the collective
fluctuations seen at long wavelength. However, even in the case that the squeezed lattice
is perfectly bipartite it appears still possible to end up with a quantum
disordered stripe anti-ferromagnet. After this detour we are back at the `mean-field
theory of sublattice bipartiteness'.

There is no reason
to assume that the strength of the exchange bonds in squeezed space is the same
everywhere. The exchange interaction ($J'$) between the spins on opposite sides of
the stripe is caused by a microscopic dynamics (hole motions) of an entirely
different kind than the superexchange which is responsible for the spin-spin
interactions inside the magnetic domains ($J$). Although nothing is known for
certain, it is generally expected that $J' < J$. If $J'$ would vanish, the
2+1D spin system associated with the ordered stripes would be cut into
independent  1+1D spin {\em ladders} with  an effective width  set by the stripe separation and
the details of the stripe ordering. If these ladders have an even width, it is 
well understood  (Dagotto and Rice, 1996) that for $J'$ is zero the system would be characterized by a spin
gap. This spin gap offers protection for the quantum disordered ladder state 
 at finite $J'$: a critical
value of $J'_c$ has to be exceeded before classical N\'eel order can emerge.
For domains of uneven width, and the case of uneven spin, the spin system on
every ladder is a Luttinger liquid (Dagotto and Rice, 1996) and any $J'$ will suffice to cause long
range order. 

This ladder notion acquires an additional significance in the light of recent
theoretical works addressing the microscopic mechanism of stripe formation.
Much of the earlier work was based, implicitly or explicitly, on the large
$S$ limit. Here the quantum spin fluctuations are neglected completely and
one finds the site ordered stripes as they first appeared in the Hartree-Fock
calculations by Zaanen and Gunnarsson (Zaanen and Gunnarsson, 1989), and variations thereof. Recently, 
Voita {\em et al.} (Voita and Sachdev, 1999; Voita, Zhang and Sachdev, 2000)
considered a limit which is in a sense opposite to large
$S$: $t-J$ type models are characterized by a global $SU(2)$ symmetry and
these can be generalized  to  a $Spl(2N)$ symmetry. By sending $N \rightarrow
\infty$, keeping $S$ finite, saddle points can be identified characterized by
exceedingly strong quantum-spin fluctuations when viewed from the large
$S$ side. In this limit, the `spin' system is generically unstable towards
the formation of spin-Peierls (or `valence bond') phases 
(Read and Sachdev, 1989; Sachdev, 1999, 2000). Nearest-neighbor
spins form pair wise singlets and these singlets are stacked in ladder-like
patterns on the 2D planes. Voita {\em et al.} showed that the large $N$
saddle-points also correspond with stripe phases as long as the hole density is not too large.  
The magnetic domains appear as Peierls-ordered even-leg ladders while the stripes
are like highly doped two-leg ladders (`bond ordered stripes'). The additional
benefit is that at large $N$, these charge stripes are generically superconducting
while uniform d-wave superconductivity takes over at large dopings. 

The most trustworthy microscopic calculations available at present are the 
numerical DMRG studies by White and Scalapino (White and Scalapino, 1998). These calculations indicate
that the stripes of the $t-J$ model are somewhere in the middle of large $S$ and large $N$. 
On the one hand, the stripes are bond centered and a case is made on basis of the 
numerics that these stripes have a tendency to become superconducting. On the
other hand, diagonal site centered filled stripes are nearby in energy and these
are quite like the Hartree-Fock stripes. In addition, it appears to be easy to stabilize 
N\'eel order and this order is characterized by a strong anti-phase boundarieness.

To obtain a better understanding of the long wavelength quantum magnetism of the stripes we studied
ourselves in great detail ladder-like spin models numerically using quantum 
Monte-Carlo (Tworzydlo {\em et al.}, 1999). These can also be considered as being representative for the 
generic spin-dynamics of the two dimensional system where the Ogata-Shiba squeezing
applies literally while the charge stripes are static. Define a $S=1/2$
nearest-neighbor Heisenberg Hamiltonian on the `squeezed' bipartite lattice.
On this lattice the stripes corresponds with a regular array in, say, the x direction
of lines extending along the y-direction  which are centered on the links of the lattice. The
exchange interaction is $J$ everywhere, except for the links which are `cut' by the
stripes where the exchange interaction is $J'$. Besides temperature, the free parameters
are (a) $\alpha = J' / J$, the ratio of the stripe-mediated exchange interaction and the
superexchange, and (b) the number of sites $N_{legs}$ separating the $J'$ links in the y 
direction: the stripe separation.

\begin{figure}
\begin{center}
\epsfig{figure=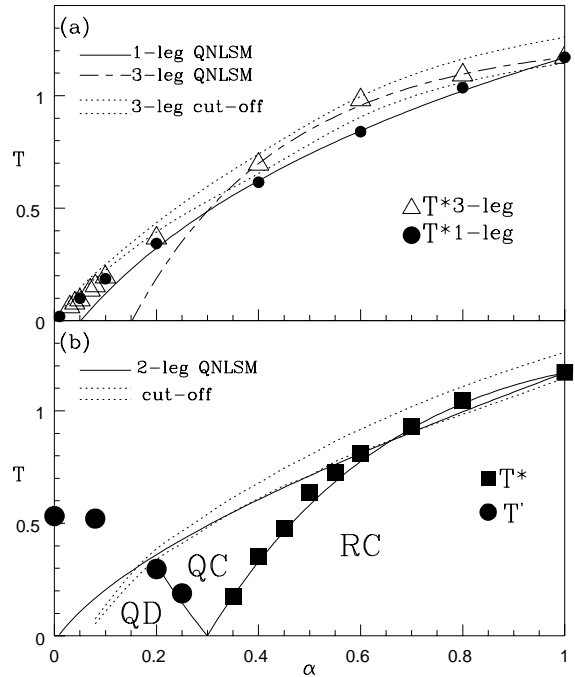, width=10cm}
\end{center}
\caption{Crossover temperatures in units of $J$ (see text) as a function of the
anisotropy $\alpha$ for the coupled one- and three leg (a), as well as the two leg (b) 
spin-ladder models. The lines and points refer to the analytical-
and numerical results, respectively, for the various scales.}
\label{yak_fig2}
\end{figure}

This is a  simple bond-dilution Heisenberg model which can be studied to any desired accuracy
using the novel cluster-loop algorithm quantum Monte-Carlo method. Obviously, for $\alpha \geq 1$
N\'eel order cannot be avoided at zero-temperature and the interest is in what happens for 
small $\alpha$. Let us therefore first consider $\alpha \rightarrow 0$. The spin system is
qualitatively different in this limit for even and uneven $N_{legs}$. For uneven $N_{legs}$
it corresponds with a disconnected system of spin-ladders with an uneven number of legs and
it is well known that these ladders renormalize in Luttinger liquids. The ladder-to-ladder
coupling $J'$ is in this situation always relevant and for any finite $\alpha$ the ground
state will exhibit long range order. The way this ground state is approached as
function of temperature is remarkably simple and is illustrated in Fig.
(\ref{yak_fig2}a) for the one-leg
and three-leg cases which appear to behave in a near-identical way. At finite temperature
all correlations are short ranged but several correlation lengths and associated characteristic
temperatures can be identified. First, as function of decreasing temperature the correlation
length $\xi_1$ in the ladder direction will rapidly increase and it is for small $\alpha$ by far the
largest length. At distances less than $\xi_1$ the spin-correlators in this direction will exhibit
the algebraic correlations of the zero temperature case. For small but finite $\alpha$ there will
be a temperature $T_0$ where the 1D correlations become so strong that even a small $\alpha$ will
suffice to cause the spins to correlate in a 2D fashion. This dimensional crossover temperature is defined
by the temperature where the correlation-length in the x direction becomes of order of twice the 
width of the ladder. We found that $T_0 \sim \alpha$ for small $\alpha$ which appears to be consistent 
with the scaling theory of Affleck and Halperin, 1996. 

At lower temperatures the spin-system is 2+1 dimensional and its long wavelength dynamics should be
characterized by the universal behaviors which follow from the $O(3)$ QNLS model. At a given temperature
$T < T_0$ there are three possibilities: \\
(a) the spin correlation length tends to saturate to a finite
value at zero-temperature. This signals that the bare coupling constant $g_0 > g^c_0$, and the system
flows to the quantum disordered zero temperature state.\\ 
(b) The correlation length behaves in the
renormalized classical fashion, meaning that the spin system is undergoing only thermal 
fluctuations although the spin-stiffness is smaller than expected because of the influence of the
quantum fluctuations at shorter scales. In this case one expects for the temperature dependence
of the correlation length $\xi (T) = \exp ( T^* / T ) / ( 2T^* + T)$ where $T^* = 2 \pi \rho_s$ in
terms of the renormalized stiffness $\rho_s$. \\
(c) The temperature is larger than either the zero temperature spin-gap associated with case (a) or
the spin-stiffness of case (b) so that the 2+1D spin system still resides in the quantum critical regime.
In this case $\xi \sim 1/T$, revealing that temperature sets the scale.       

As it turns out, in the uneven-leg cases the system jumps directly into the renormalized classical regime
at the moment it finds out that it it becomes two dimensional. Hence, at higher temperatures the physics
is that of decoupled 1+1D subsystems and these discover that they are on the way to a classical 2D 
N\'eel state at the moment that the temperature is low enough such that these 1+1D subsystems start
to correlate in a 2+1D fashion. It is even so that the renormalized spin stiffness $T^* \simeq T^0$,
the dimensional crossover temperature.

The even-leg case is far more interesting and this is illustrated in
 Fig.\ref{yak_fig2}b for the two-leg ladders.
Even leg ladders in isolation ($\alpha=0$) are well known to exhibit a spin gap. This phenomenon is
probably best understood as a consequence of the fact that the Peierls states are the natural competitors
of N\'eel order for lattice-quantum antiferromagnets (Read and Sachdev, 1989; Sachdev, 1999, 2000). 
Define a Peierls order parameter field which amounts
in the two leg ladder case to stronger exchange bonds along the rungs of the ladder and weaker bonds
along the legs. Obviously, if this difference is large enough two spin singlets are formed along the rungs, and
the overall state of the system can be viewed as a simple row of these rung-singlets and the system has a
spin gap. Upon reducing this difference to zero, the spin gap stays finite and therefore the state of the
spin ladder with uniform exchange couplings is adiabatically connected to the Peierl's state where translational
symmetry is explicitly broken. For four leg-, six leg ladder, etcetera,  cases the same argument can be used, except that
one is now dealing with 2 and 3 parallel rows of singlets, respectively. The spin gap decreases rapidly
with the number of legs in the absence of frustrations. However, adding frustrations it appears that a stable
phase exists where this gap stays open even if the number of legs approaches infinity (Read and Sachdev, 1989;
du Croo-de Jongh, van Leeuwen and van Saarloos, 2000). It is noted that this
logic does not apply to uneven leg ladders because of the relevancy of the topological phases associated with
the uneveness of the total spin on every rung.

Let us now consider the cross-over diagram as function of $\alpha$ and temperature for the coupled two-leg ladder
problem. For $\alpha$ of order 1 the spin system is just like the uniform Heisenberg problem on a square
lattice. At a temperature $T/J \simeq 1$ the correlation length is of order the lattice constant and from Fig.
\ref{yak_fig2}b one infers the well known result that this Heisenberg problem (of relevance to the half-filled case) is
so far away from the critical point that before the crossover to the quantum critical point is reached the
correlation length has already hit the lattice constant: at all temperatures this system is in the renormalized
classical regime. This changes when $\alpha$ is reduced. As in the uneven leg cases, the dimensional
cross-over temperature $T_0$ decreases well. However, the renormalized stiffness of the 2D renormalized
classical regime decreases more rapidly and a window opens up of genuine 2+1 dimensional quantum criticality. This
is of course due to the quantum phase transition to the Peierl's state occurring at $\alpha_c \sim 0.3$. It is
noticed that for the two leg ladder magnetic domains this transition occurs while $J'$ is still quite
substantial. In the proximity of the quantum-critical point one is dealing with the competition between the
N\'eel state and a truely 2+1 dimensional spin-Peierls instability which is just helped by a partial explicit breaking
of translational symmetry. It is only for $\alpha < 0.2$ that the one dimensional on-ladder spin dynamics gets
protected: the on-ladder correlation length never grows large enough to fulfill the conditions needed to cause
a 2+1D spin dynamics, and in this regime one is dealing with decoupled ladders. It is noticed that all these
behaviors have been reproduced in quite some detail using the scaling theory for the spatially anisotropic
$O(3)$ quantum non-linear sigma model by van Duin and Zaanen, 1998, extending earlier work by Castro-Neto and Hone, 1996.

\section{Unifying spins and stripes: a simulation.}

\begin{figure}[t]
\begin{center}
\epsfig{file=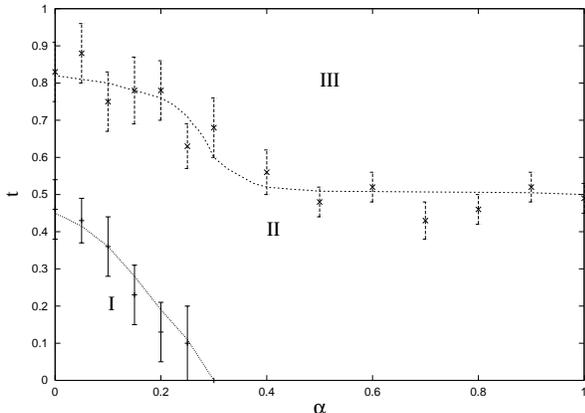,width=8cm}
\caption{Phase diagram of the model describing quantum stripes fluctuating through a quantum
anti-ferromagnet, for the particular case that stripe are separated at average by two sites.
$\alpha$ is the spin-anisotropy parameters and $t$ is the stripe kinetic energy, both
measured in units of $J$.
Phase I corresponds with a charge-ordered and spin quantum disordered stripes, while regions
II and III correspond with full charge and antiferromagnetic order. However, regions II and III are
separated by a sharp crossover associated with the single stripe unbinding transition.}
\label{Phasediag}
\end{center}
\end{figure}

Starting out with the Ogata-Shiba principle of section II, we deduced a general problem and
subsequently we studied several aspects of this problem separately. Of course, the whole can
be more than the sum of its parts. What happens when the lattice strings of section IV form the
interacting string system of section V, which is in turn communicating with the quantum-antiferromagnet
of section VI? The answer is the phase diagram shown in Fig. \ref{Phasediag} (Osman, 2000). The control parameters are the
$\alpha$ of the previous section parametrizing the relative strength of the stripe-mediated spin-spin
coupling and the hopping parameter $t$ of section IV, controlling the strength of the stripe quantum fluctuations.
Finally, the stripe density matters and the phase diagram is the one for a density such that the
domains form two-leg ladders when the stripes are static. We also looked at different densities.
The topology of the phase diagram is the same for all commensurate stripe densities such that
the magnetic domains form even-leg ladders. For increasing ladder width the parameter regime
where phase I is stable shrinks rapidly, while it disappears completely for all other densities.
In all other regards the resulting phase diagram is generic.

There are actually only two phases: the charge-ordered incompressible stripe antiferromagnet found for
small $\alpha$ and $t$ (phase I) and a fully charge and spin ordered phase (II and III). However,
although `phases' II and III are indistinguishable at 
long wavelength their short distance nature is very different. The line
separating these two phases is no more than a cross-over line but it is a very sharp cross-over
which was actually detected in the numerics using the Binder criterion (Binder, 1997) which is devised for
phase transitions. This cross-over line corresponds with the single string unbinding transition
of section IV. Hence, in phase II a single stripe breaks already translational symmetry and the
2D order in the localized stripe system  is driven by stripe-stripe interactions mediated by
the spin system. In phase III every individual stripe is delocalized but the order-out-of-disorder
mechanism as discussed in section V takes over, causing again translational symmetry breaking.
However, this is a very weak order. Although the numerical calculations confirmed this mechanism
it is in many cases very hard to detect. The novelty is that the interplay of the quantum spin dynamics of section
VI and the stripe fluctuations gives rise to a next surprise. {\em Sufficiently strong stripe-quantum
fluctuations restore the N\'eel order even if the spin system of the static stripes would be quantum
disordered.} Although this N\'eel order is weak when measured by the standard spin-correlator,  the
hidden spin-order as measured by the topological correlator Eq. (\ref{topcor}) as discussed in section II 
approaches the magnitude of order as established in  the Heisenberg system of the half-filled insulator,
but only so when $t$ is large. The conclusion is that by just imposing the Ogata-Shiba squeezing condition, 
quantum fluctuations turn into agents, mediating every order which can be realized in this system. 

\begin{figure*}[t!]
\centering
\epsfig{file=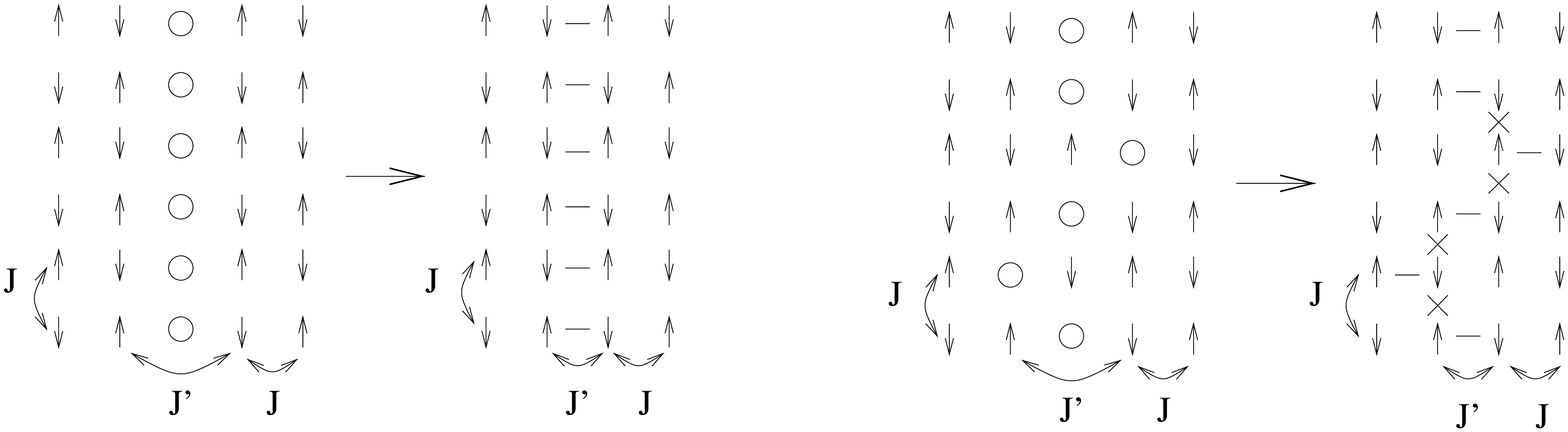,width=14cm}
\caption{ The model  describing  fluctuating stripes in the quantum antiferromagnet. Starting with various connected
stripe configurations, one arrives at a spin-only model in squeezed space, characterized by fluctuating lines of
modified exchange bonds. Notice that in squeezed space exchange bonds disappear (crosses) when the original stripe
configuration contains diagonal pieces.}
\label{fig:cha51}
\end{figure*}

Let us discuss in more detail the model and the numerical simulations
behind this phase diagram. The model is most easily constructed in squeezed space. Hence, we consider 
a two dimensional square lattice with a Heisenberg $S=1/2$ spin on every site. These spins interact
via nearest-neighbor interactions $J = 1$ except along 1+1D connected trajectories defined on the
links between the sites which are characterized by a weaker exchange interaction $J' = \alpha J$,
see Fig. (\ref{fig:cha51}).
Different from the model in Section V these trajectories can now have arbitrary shapes (see Fig. \ref{fig:cha51}), 
and by unsqueezing the lattice one recovers the lattice strings. Notice that
the connectedness constraint that the holes have to 
be nearest- or next-nearest-neighbors translates
on the squeezed lattice in the constraint that the $J'$ links have to be nearest- or next-nearest-neighbors
on the {\em link} lattice. Notice also that for diagonal stripe configurations,  nearest-neighbor spin
bonds emerge on the squeezed lattice, corresponding  with next-nearest-neighbor spins on the unsqueezed lattice 
(crosses in Fig. \ref{fig:cha51})
 and these particular squeezed lattice links carry therefore no spin-spin interaction. As we
will discuss later in more detail, these missing exchange interactions are equivalent to the parameter $K$
in the string model of Section IV, expressing the energy difference between nearest- and next-nearest-neighbor hole bonds.
All other curvature parameters (the $L$'s) are set to zero.

The $J'$ strings are quantized in the same way as the strings of section IV. The $J'$
links can hop to nearest-neighbor positions on the link lattice as long as these moves do not
violate the connectedness constraint. We learned in section III that this combination always leads
to the directedness symmetry breaking. Hence, undirected string configurations do not have to be taken
into account, which simplifies the calculations considerably. At the same time this means that 
dislocations cannot occur which represents a serious limitation. However, this limitation follows 
from the `mean-field' requirement that the squeezed lattice has to be partite everywhere.

Finally, we impose a hard-condition by requiring that $J'$ bonds cannot meet on the same link of the
same lattice. As can be easily checked, this means in the unsqueezed space that two stripes should be
separated by at least one spin site. The hard-core is therefore spread out and this does not matter,
although one should be aware of it when the stripe density is high. 

This defines the model and let us for completeness formulate the Hamiltonian explicitly. Define 
coordinates (x,y) for the sites on the squeezed lattice. Consider the stripes to be directed along the
$y$ direction. Define hard-core particles $a^{\dagger}_{(x,y)}$ ($n_{(x,y)} = a^{\dagger}_{(x,y)} a_{(x,y)}$)
which live on the links of the squeezed
lattice where $(x,y)$ labels the link connecting the  site $(x,y)$ with the site $(x+1,y)$. The
Hamiltonian is,
\begin{eqnarray}
H & = & t \sum_{x,y} (  {\cal P} ( a^{\dagger}_{(x+1,y} a_{(x,y)} + h.c. ) {\cal P} \nonumber \\  
 &  & + J ( 1 - (1 - \alpha) n_{(x,y)}) \vec{S}_{(x,y)} \cdot \vec{S}_{(x+1,y)} \nonumber \\
 &  & + J ( 1 - ( n_{(x,y)} n_{(x-1,y+1)} \nonumber \\
 &  &  + n_{(x-1,y)} n_{(x,y+1)} ) \vec{S}_{(x,y)} \cdot \vec{S}_{(x,y+1)} )
\label{fullmod}
\end{eqnarray}
The first term corresponds with the stripe kinetic energy where ${\cal P}$ is the projection operator
of Eq. (\ref{strproj1}), but now defined on the links,  ensuring that hole hoppings do not break up the strings.
The second term changes the exchange $J$ into $J'$ when the particle is present on the bond,
and the last term takes care of the missing exchange interactions associated with the diagonal stripes
(the crosses in Fig. \ref{fig:cha51}).

\begin{figure*}[t!]
\centering
\epsfig{file=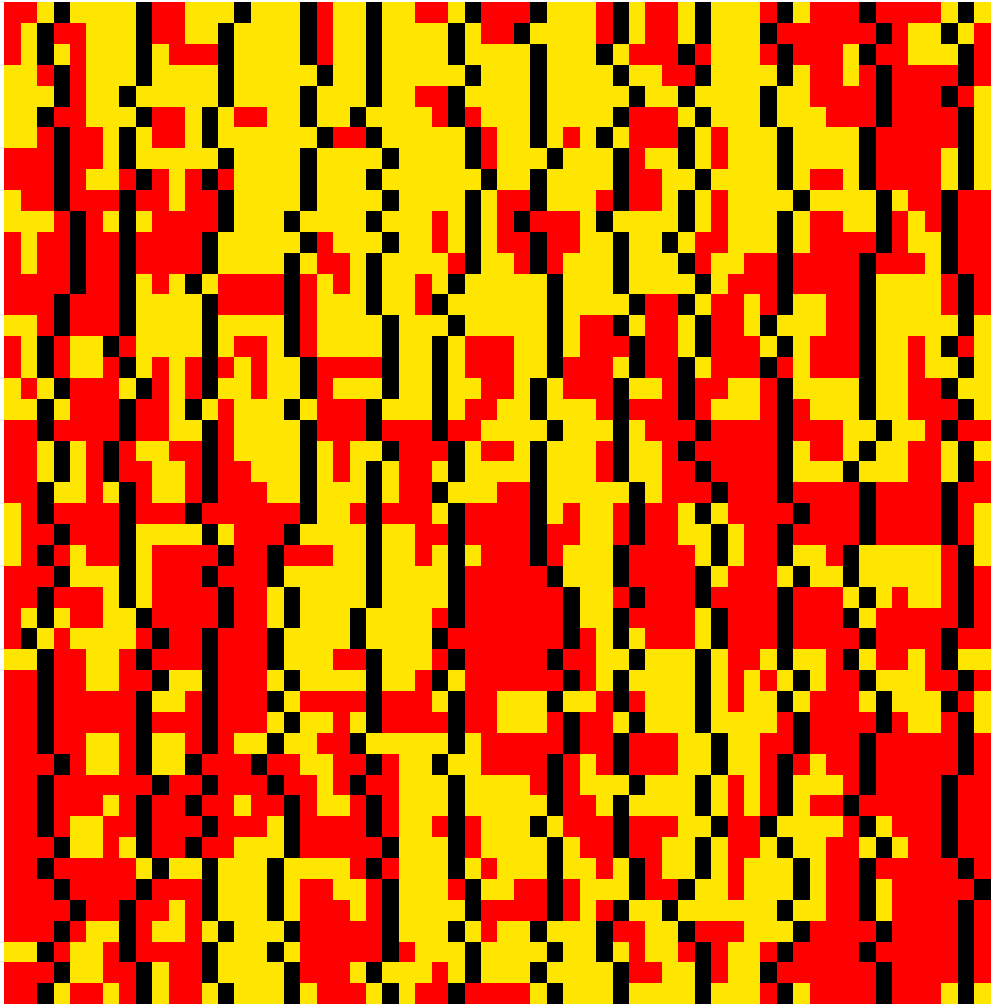,width=14cm}
\caption{ A snapshot of a timeslice of the quantum Monte-Carlo simulation for a system of 12 stripes which
are at average four sites apart.
The black lines are the stripes and the light- and middle gray shades
(yellow and red
in the color version) indicate opposite orientations of the hidden
spin-order parameter
(staggered spin on the squeezed lattice).}
\label{timeslice}
\end{figure*}

We studied this model numerically using a `hybrid' Monte-Carlo algorithm for the spins and the strings.
The updating can be done quite independently in these subsectors because strings and spins live on different
(site vs. link) lattices. The spin system was simulated using the loop-cluster algorithm (Evertz, Lana and Marcu, 1993)
and improved estimators (Wiese and Ying, 1994) which were also
used in the simulations discussed in section VI. The string system was simulated using the same world-line algorithm
as employed for the single string problem, which is described at length in Eskes {\em et al.}, 1998. Although
this latter algorithm is a conventional Monte-Carlo algorithm, and therefore not as efficient as the
loop-cluster algorithm used for the spin system, the dynamics is relatively simple and we could simulate
systems of up to 20  strings on a squeezed lattice of $60 \times 60$ sites. To give some idea, in
Fig. \ref{timeslice} a snap-shot is shown of a time-slice for a system of 12 strings which are at average separated
by four-leg ladder domains, represented in unsqueezed space.

To determine the phase boundaries and cross-over lines in the phase diagram 
Fig. \ref{Phasediag} more accurately we
 used the Binder parameter technique (Binder, 1997). 
This parameter is the reduced fourth order cummulant defined as
\begin{equation} 
  \label{eq:cha5_2} 
   B = 1 - {\langle O^{4} \rangle \over 3\langle O^{2}{\rangle}^{2}}
\end{equation}
where $O$ is the order parameter of the ordered phase. This parameter behaves differently in the ordered
and disordered state. In the ordered state $\langle O^4 \rangle = \langle O^2 \rangle^2$
and therefore B = 2/3 while in the disordered state characterized by Gaussian fluctuations    
$\langle O^4 \rangle = 3 \langle O^2 \rangle^2$ and therefore $B = 0$. This technique also turned out
to be useful for finding the cross-over line between regimes II and III, indicating that the 
order parameter is suddenly strongly reduced at this cross-over.

Let us now turn to the phase-diagram. The case $t=0$ just corresponds with the spin-only physics discussed
in the previous section and the issue is what happens at finite $t$. A particular simple case is where 
$\alpha = 1$. The spins and strings move independently except for the last term in Eq. (\ref{fullmod}).
We already indicated that this term is equivalent to a finite ${\cal K}$ parameter in the string-only
problem. A positive and sufficiently large ${\cal K}$ localizes a single string along the `vertical direction'
and only when $t$ exceeds a critical value the string unbinds from the lattice. This single string unbinding
transition is responsible for the sharp cross-over between the II and III regimes in the diagram Fig. \ref{Phasediag}.
The critical $t$ can be easily estimated. A kink in the string will break a spin-bond along the y-axis and this
will cost an energy equal to the energy per bond in the pure Heisenberg spin system. The energy per site
has been calculated by many groups to be $-0.6692 J$. Given that for every site there are two bonds, and 
given that the cost of a kink ${\cal K}$ is equal to that of a spin bond, we find ${\cal K} = 0.34 J$. Eskes
{\em et al}. found that the transition from the flat- to Gaussian string phases occurs when ${\cal K} = 0.7 t_c$
and we therefore estimate the locus of the cross-over to be at $t_c = 0.34J/ 0.7 = 0.49 J$, in striking agreement 
with the simulations.

\begin{figure}
\begin{center}
\epsfig{file=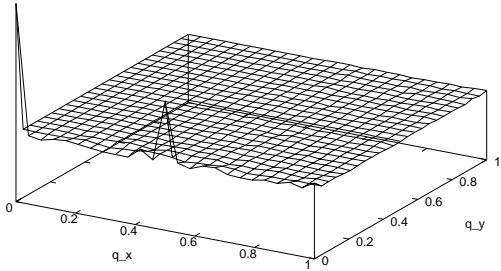,width=6cm,angle=270}    
\caption{Plot of the charge-charge structural factor for a lattice of 
$48 \times 48$ at the point $\alpha = 0.05$ and $t = 8.0$ inside phase III. 
The average distance between the strings is 4. }
\label{StrFac}
\end{center}
\end{figure}

For $t < t_c$ there is definitely a very strong stripe ordering tendency. Every stripe is like a rigid rod, and
any stripe-stripe interaction will lead to translational symmetry breaking towards a periodic state. We will
in a moment discuss the nature of the stripe-stripe interactions mediated by the spin system. This is different
in the regime $t > t_c$. The stripes by themselves delocalize and the ordering tendency is now driven by the
order-out-of-disorder effects discussed in section V. This is a very weak order and we actually did not manage
to detect it in the simulations for densities where the stripes are separated by two-leg ladders at average. This
in fact illustrates vividly the order-out-of-disorder mechanism. Because of the `smeared out' hard-core of the
stripes, the system is very dense at this stripe separation and the fluctuations are therefore strongly reduced.
Intuitively one would expect that there would be even less tendency towards order if the stripes are placed 
further apart, but the opposite happens. We calculated the charge-structure factor on the unsqueezed lattice
for a system of 12 stripes which would have a periodicity in the ordered state of 5 lattice constants (domains
are 4 legs wide, Fig. \ref{timeslice}) at a large $t = 8J$. 
As can be seen from Fig. \ref{StrFac}, this structure factor is characterized
by a quite sizable `$2 \varepsilon$' charge order peak, relative to the lattice Bragg peak located at the 
origin. Although this represents a striking qualitative confirmation of  the string-gas order-out-of-disorder mechanism
(Zaanen, 2000) we did not attempt to further quantify these matters because of the rather serious size limitations of our
simulations.

Let us now turn to the interesting aspects of the interplay between the spin- and the string system which 
emerges when $\alpha$ becomes small. A first issue is that the spin system mediates interactions between the stripes,
even when the stripes are static (small $t$). This has been studied in quite some detail by Pryadko, Kivelson and Hone, 1998. 
Assuming small $\alpha$, the stripes correspond with reflecting boundaries for the spin
waves and as a result attractive Casimir forces arise which would render the stripe system unstable 
towards phase separation in the absence of compensating repulsive long range forces (like, e.g., 
direct Coulomb interactions). This Casimir potential falls off like $V (d) \sim 1/d$ where $d$ is
the stripe separation. However, this long-wavelength analysis is not quite applicable to the 
most relevant cases where stripes are only a few lattice constants apart. In this situation, the ladder effects
as discussed in the previous section are expected to become also quite important with regard to
the stripe-stripe interactions. We studied this problem by inserting two static horizontal stripes
separated by $d$ spin sites on a large Heisenberg lattice, comparing its
energy with that of a single stripe, and the energy of the pure spin system. The stripe-stripe interaction energy
per unit length of stripe is by definition,
\begin{equation}
V_{int} ( d ) = ( E_2 (d) + E_0 (d) - 2 E_1 (d) ) / N_{l}
\label{inststr}
\end{equation}
where $E_n$ is the total energy of the system with $n$ stripes while $N_l$ is the stripe length.
We have calculated $V_{int}$ for both $\alpha =0.2$ and $\alpha = 0.08$ and the results are
shown in Fig. 11.
 The major surprise is that even for stripe separations as small as two lattice
constants these interactions are very weak, $\simeq 0.02J$, to become even more minute
at larger distances. The two leg ladder case appears to be exceptionally stable, while the
calculations suggest that even separations are always more stable than the uneven cases,
as expected from the presence of a spin gap in the former. For instance, if all other forces
could be neglected, these spin-mediated interactions would render a stripe system with $d=3$
to become unstable to a stripe `density wave' characterized by $d=2, 4$: $\cdots - 3 - 3 - 3 - 3 - \cdots
\rightarrow  \cdots - 2 - 4 - 2 - 4 - \cdots$. However, it appears that in reality these quite
feeble forces would be easily overwhelmed by interactions from other sources, like electron-phonon
coupling and direct Coulomb interactions.
\begin{figure}[h]
\begin{center}
\epsfig{file=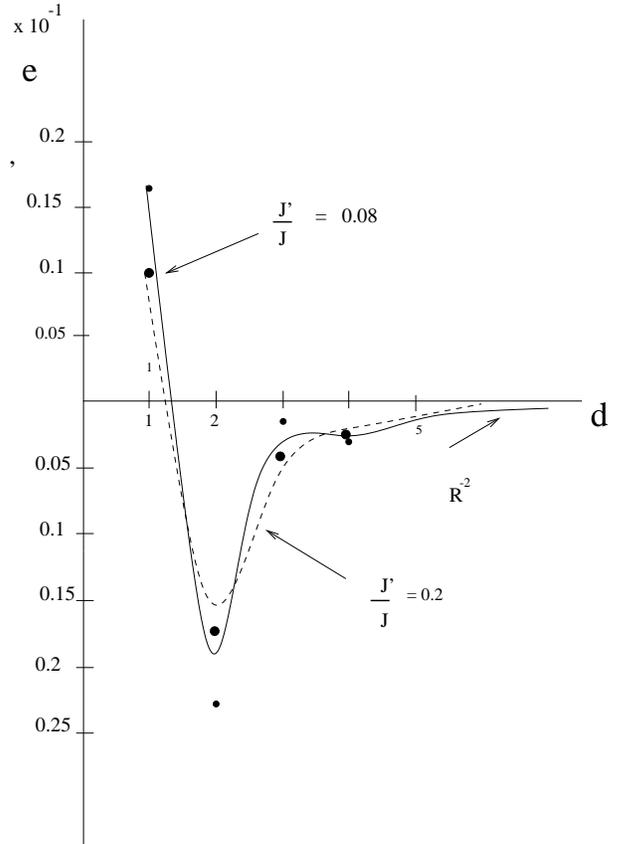,width=8cm}  
      
\caption{Induced interaction between two static flat stripes. $d$ is the 
distance between the two stripes and $e$ is the energy per unit length of 
stripe.}
\end{center}
\label{fig:casimir}
\end{figure}

\begin{figure}[t]
  \begin{center}
\epsfig{file=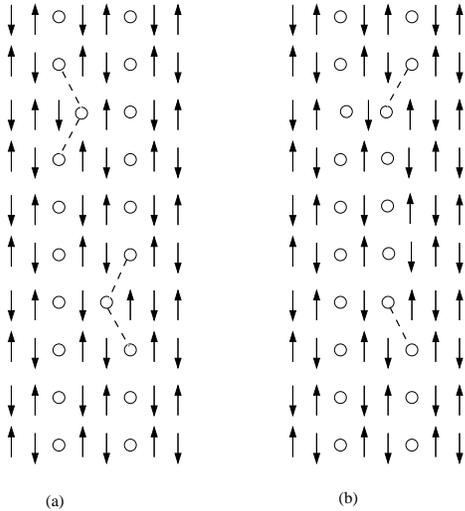,width=6cm}    
  \end{center}
\caption{(a) Bound kink-antikink pairs in a coupled two-leg ladders stripe system. (b) When the 
kinks unbind strings of odd-leg ladders develop corresponding with a confining potential. It is
believed that this mechanism is responsible for the upturn of the single string unbinding cross-over
at small $\alpha$.}
    \label{linpo}
\end{figure}

We discussed in the previous section that static stripes separated by even leg ladders give rise
to a quantum disordered phase for small $\alpha$. An interesting result is that strong stripe
fluctuations restore the N\'eel state, and the gap of the incompressible  spin phase (phase I)
vanishes even for the two leg-ladder case before the single stripes unbind from the lattice
(see Fig. \ref{Phasediag}). In hindsight this is not so surprising. The `hole' motions will cause kinks in the
stripes and this means that the two leg-ladder is locally destroyed: over some length the spin domain
becomes a 1 leg or 3 leg ladder (Fig. \ref{linpo}). In this regard, it is interesting that the single stripe unbinding
crossover increases substantially at small $\alpha$'s, more or less tracking the magnitude of the
spin-gap in the incompressible regime. Apparently this crossover scale is no longer completely due
to the simple missing exchange bond mechanism we discussed earlier. We also learned that the energy
associated with changing the stripe separation locally is quite small (previous paragraph) and 
actually too small to explain the upturn of the cross-over line. Therefore, this upturn is caused
by some non-trivial mechanism. We suspect that 
this is of the kind as illustrated in Fig. \ref{linpo}b. In order
for a single stripe to unbind, kinks should deconfine: isolated  side steps of the strings should 
proliferate freely in the vacuum. From Fig. \ref{linpo} one infers that for this too happen, strings of
one-leg and three-leg ladders have to be created and these exert clearly a confining force on the
kinks and such a confining force is not considered in the simple string model of Eskes {\em et al.}, 1996, 1998 .

With regard to the recurrence of N\'eel order, this destruction of two-leg ladderness is not all.
In close analogy with the 1+1D case, at large $t$ the holes are fluctuating rather freely and
these hole motions induce a direct spin-spin interaction:
let one hole hop back and forth and one directly infers that this causes a strong
preference for the spins on both sides of the stripes to be antiparallel. Hence, for $t$ large as
compared to $J$ these hole motions average away the difference between $J$ and $J'$ and at large
distances the exchange interactions can be taken to be uniform. Hence, $\Delta J = J - J'$ is in
this sense an irrelevant operator and one recovers a notion of spin charge separation which is
quite like the one encountered in  the 1+1D Luttinger liquid. The difference is of course that in
2+1D the spin system orders.

\begin{figure}
\begin{center}
\epsfig{file=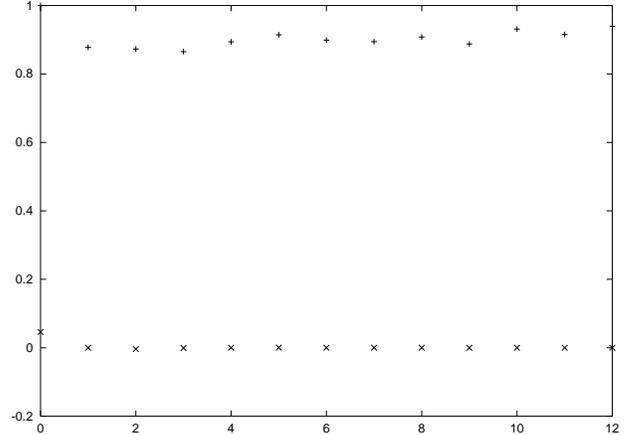,width=6cm,angle=270}    
\caption{The behavior  of the topological spin-spin correlator (pluses) and the 
direct staggered spin-spin correlator (crosses) as function of distance on a
$24 \times 24$  squeezed  lattice at a two-leg ladder stripe density ($\alpha = 0.05$, $t = 2.0$). }
\label{TaDcor}
\end{center}
\end{figure}

In squeezed space this charge-fluctuation induced N\'eel order is very easy to observe. The 
topological spin correlator, Eq. (\ref{topcor}),  which measures this order in unsqueezed space (by `dividing out'
the anti-phase boundarieness attached to the charge) is equivalent to the direct spin correlator
in squeezed space. At the same time, starting in squeezed space the direct spin correlator
as measured in unsqueezed space is easily computed by reinserting the antiphase boundaries 
using a simple algorithm. In Fig. \ref{TaDcor} we show  typical results for the direct and
topological spin correlators in unsqueezed space in region III of the phase diagram Fig. \ref{Phasediag},
 calculated for a density corresponding with two-leg spin ladders. It is
seen that the topological correlator barely decays and it behaves in the same way as the
staggered spin correlation function of the $S=1/2$ Heisenberg model on the square lattice.
At the same time, the direct staggered spin correlator does not show {\em any sign of
order}. As we already discussed, at these high stripe  densities the `string gas' induced stripe
order becomes very weak and the disorderly behavior of the direct spin correlator is entirely
due to the quantum disorder in the stripe sector. Hence, Fig. \ref{TaDcor}
 demonstrates that
despite the presence of a `hidden' spin order which is as strong as the N\'eel order
found in the pure spin problem, this can be obscured completely from the view of the
experimentalist which can only measure the direct spin correlator which is affected by
the fluctuating anti-phase boundaries. This is a lesson to keep in mind when confronted
with claims that dynamical stripes do not exist  because they do not appear in
measured dynamical spin susceptibilities.
 
\section{Destroying sublattice parity order.}

In the previous sections we have described a theory of stripe {\em order} as it emerges from 
a microscopy dominated by quantum fluctuations. In fact, when we started the research described
in the above, we were after the physics of stripe quantum disordered states. The truely interesting
problem is of course the phenomenon called `dynamical stripes', referring to experimental
anomalies found in the fully developed superconductors. If it has anything to do with stripes, it
has to be that these stripes phases are quantum disordered in an essential way. Hence, when we
discovered the spontaneous directedness of the Eskes strings, followed by the string-gas order-out-of-disorder
mechanism we were at first disappointed. It took a while to just appreciate the above on its own merit.

We expect that any reader who is aware of the relatively developed theoretical understanding of the
microscopy of the stripes  would have noticed that the `stripyness' as imposed by the perfect sublattice
parity order is too literal. Starting  with the true holes and spins as described by $t-J$ type
models one ends up with a picture which is much less orderly in this regard (Zaanen, 1998; White and Scalapino,
1998; Morais-Smith {\em et al.}, 1998; Chernyshev {\em et al.}, 2000; Tchernyshyov and Pryadko, 2000; Martin
{\em et al.}, 2000).  For realistic values of
parameters the holes are much more loosely bound to the stripes and one might even wonder if at the
densities of interest to the superconductors one can uniquely assign a particular real hole to a particular
stripe. We are well aware of this and the above should be considered as no more than a fixed-point theory,
a strong coupling limit with regard to the stripyness which will have  a finite basin of
attraction. The physics described in the above will be robust against some degree of local
violation of the order, and in this restricted sense it might tell something about the origin of
stripe order. At the same time, this robustness has its limitations and when the fluctuations increase
at some point a phase transition has to follow where sublattice parity order is destroyed. At the same
time, it could well be that the local sublattice parity fluctuations disorder charge and/or spin well
before the transition occurs where sublattice parity order vanishes.
Therefore, a variety of distinct, partially disordered phases can exist in between the
fully ordered stripe phase and the fully disordered state and these might have something to do
with high Tc superconductivity (Zaanen and Nussinov, 2000).

\begin{figure*}[t!]
\centering
\epsfig{file=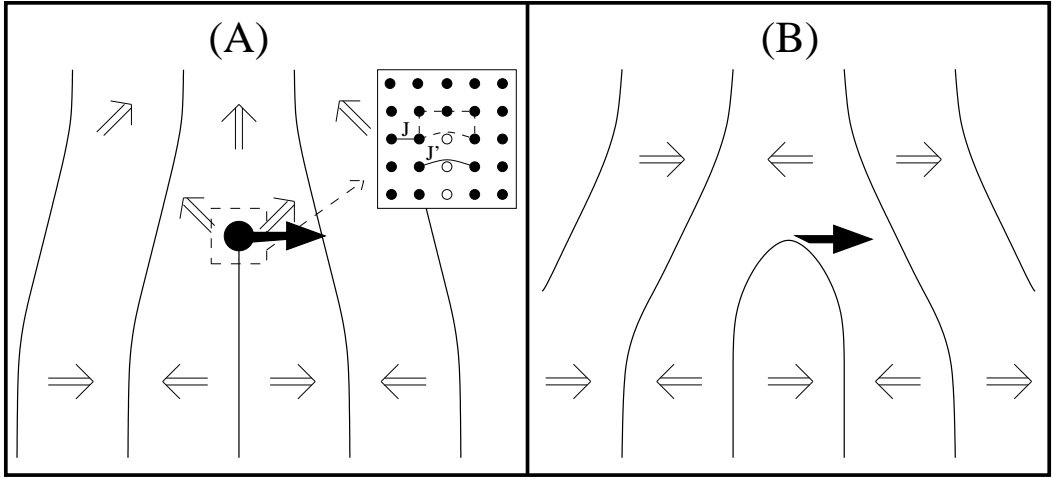,width=14cm}
\caption{ (a) The fundamental stripe dislocation which not only destroys the charge order, but also the sublattice parity
order, while it gives rise to an essential frustration in the spin system. The bold arrow corresponds with the
Burger's vector while the open arrows indicate the half-vortex like texture it causes in a semi-classical 
antiferromagnet. In the inset its action in squeezed space is indicated. (b) Initially a dislocation and an
anti-dislocation (with regard to sublattice parity) will bind due to their logarithmic spin-mediated interaction.
This sublattice parity `neutral' dislocation will still destroy the charge order although it is no longer an
essential frustration in the spin system. }
\label{dislocation}
\end{figure*}

What are the fluctuations violating sublattice parity order? As we already stated repeatedly, perfect
order of this kind means that stripes form trajectories of nearest- and next-nearest neighbor links
on the lattice and to violate this order one has to violate this connectedness requirement. The disorder
excitation is therefore simple and unique (Zaanen and Nussinov, 2000): 
a stripe coming to an end, an object which we called the `stripe dislocation'
(Fig. \ref{dislocation}).

This entity has clearly a topological status: a single stripe dislocations destroys sublattice parity 
everywhere. The sublattice parity in the region `below' the stripe dislocation cannot be matched consistently
with the parity `above' the dislocation (Fig. \ref{dislocation}) and thereby 
it destroys the notion of a definite sublattice parity.
At the same time, it is a classic dislocation with regard to the charge order. It is like the half row of atoms
of metallurgy which is well known to be the topological excitation destroying the translational symmetry breaking
associated with crystalline order. Finally, it represents clearly a disorder event with regard to N\'eel order --
it destroys the bipartiteness of the squeezed lattice and thereby it represents an essential spin frustration 
which will destroy the N\'eel order everywhere as well. However, it is not quite a genuine
topological excitation of the spin system, as we will discuss later.

Despite its simple appearance, this stripe dislocation has the remarkable meaning that it is the omnipotent
disorder excitation belonging to stripe order. It destroys at the same time anti-phase boundarieness, charge-
and spin order. Its significance becomes particularly obvious in combination with the general principle of
duality. 

Duality is a mathematical principle with a general applicability in continuum field theory, stating
that a deep relation exists between states of matter separated by a phase transition. In the condensed matter
context its applicability is limited to situations where one can stay away from the lattice (UV) cut-off, which
means in practice that it has little to say about strong first order transitions. This relation is as follows.
Given a long range order, excitations can be uniquely defined using the machinery of topology which destroy this
order globally. The field configurations of the continuum theory can be rigorously subdivided in smooth configurations
and singular configurations corresponding with the topological excitations. The smooth configurations cannot destroy
the order because they are themselves part of the order and the disorder is carried entirely by the topological
excitations. Hence, at the order-disorder transition the topological excitations proliferate in the vacuum. Because
these topological excitations are interacting entities occurring at a finite density in the disordered states they
in turn define an interacting system with a tendency to break symmetry spontaneously. The order parameter theory
of this disorder matter is than equal to the field theory describing the long-wavelength physics of the disordered
state. We are intrigued by the following question: {\em could it be that static stripes and the high Tc superconductors
are related by duality ?} If this would be the case the physics of the superconductors should be related to the 
physics of stripe dislocation matter, because the stripe dislocations are the elementary disorder excitations 
associated with stripe order. Different from the theory of stripe order, we do not understand the nature
of stripe disorder theory at all.

The fundamental problem is that at least for spin $S=1/2$ the stripe
dislocations are essential frustrations (Zaanen and Nussinov, 2000) restoring a Marshall sign
structure in the vacuum, and these signs represent a difficult but not
necessarily intractable problem (Weng {\em et al.}, 1997). Without these signs we have some
understanding of the disorder theory and let us sketch some of the
essentials. Let us assume that the charge system is made out of bosons
while the spin system can be represented with the $O(3)$ quantum non
linear sigma model, which is also a bosonic field theory.  Except that
we now allow for stripe dislocations, and the neglect of the Marshall
signs of the spins, everything else is like the situation described in
Section VII. What can happen?  The answer is that the fully ordered
stripe phase melts initially in a charge quantum nematic which is at
the same time a {\em spin nematic}. Subsequently, this state can
either undergo a first order transition into an isotropic quantum spin
liquid, or in a incompressible quantum spin liquid characterized by a
topological order.  The effective field theory governing this spin
dynamics is the $Z_2$ gauged $O(3)$ quantum non-linear sigma model.

The qualitative idea is straightforward. Consider the stripe dislocation, insisting that the spin system is semiclassical.
It is directly clear that the $O(3)$ order parameter will fold around the dislocation forming a half-vortex like
texture, see Fig. \ref{dislocation}. 
Interestingly, this is not a topological excitation in the $O(3)$ spin system. The topological excitation
associated with the internal $O(3)$ symmetry in 2+1 D is a skyrmion (Fradkin, 1991), 
and this is a texture living on the time slice
which involves rotations in order parameter space  in two orthogonal directions. The half vortex associated with the stripe dislocation 
rotates only in one plane and it is easily seen that it carries in addition a zero-mode associated with the rotation in the 
orthogonal plane. Despite these intricacies, it is still true that on the time slice these half-vortices interact via long
range, logarithmic interactions mediated by the spin system. In this sense they are like $O(2)$ vortices and these
logarithmic interactions will cause dislocations and antidislocations to bind in pairs initially, which are globally equivalent to
the `neutral' dislocation  indicated in Fig. \ref{dislocation}. This bound pair of dislocations is not affecting the sublattice 
parity globally. At the same time
 it is a dislocation with regard to the charge system 
carrying a Burger's vector which
is twice that of the elementary dislocation. 

Referring back to section V, the order-out-of-disorder argument was 
based on the observation that dislocations of the type Fig. \ref{dislocation}
cannot proliferate. However, this could only be proved under  assumption that stripes are uninterrupted elastic lines. 
Allowing  for a small
but finite break up probability invalidates this assumption and thereby the argument. End points of stripes 
fluctuate much more than intact stripes and therefore they will proliferate always when the kinetic energy is
large enough. However, because of the argument presented in the previous paragraph,  sublattice parity order is still protected
because of  the logarithmic interactions mediated  by the spin system.

Since these neutral dislocations restore translational invariance, 
charge order is destroyed and the stripes form a quantum fluid. In fact,
this is a nematic quantum liquid crystal, of the kind introduced by Kivelson, Fradkin and Emery, 1998.
However, this is a subject of its own which is not essential to the remainder of
the present argument. What matters is that the charge disorder implies a gap 
and the next question is on the nature of the spin dynamics at energies less
than this mass-gap.

Since the neutral dislocations do not destroy sublattice parity globally,   the
`hidden' spin order of section VII can survive in principle. These dislocations 
cause local frustrations which will act to decrease the hidden N\'eel order but this does not necessarily imply
that the order completely disappears at the moment that these dislocations  proliferate. Hence, in principle
a state exists characterized by the hidden N\'eel order as measured by the topological correlator
Eq. (\ref{topcor}), while the charge sector is quantum disordered. This state is {\em a spin nematic}.

Spin nematics were first introduced by Andreev and Grishchuk, 1984, on basis of general symmetry considerations.
However, to the best of our knowledge these spin nematics  have never been identified in experiment, and we are claiming
that this type of spin order arises in a most natural way in this stripe context. Consider a snapshot
of a time-slice in Euclidean space time, of the kind as shown in Fig. 
\ref{timeslice}. Point the finger at a particular site in a
magnetic domain. Because the dislocations are proliferated, stripes are delocalized and it has to be that
in going along the time direction at some point a stripe will pass this particular site and after this
passage the A sublattice has changed in the B sublattice and vice versa. Therefore the N\'eel order parameter
will point in exactly the opposite direction. There is still a sense of broken spin rotational symmetry. Call the initial
direction the north pole. After the stripe has passed the order parameter will point to the south pole. Hence,
north pole and south pole are identified but the location of the north pole can still be freely chosen on
the half-sphere. This is the director order parameter which is usually associated with nematic order, and therefore
this state should be called a spin nematic.

Traditionally (de Gennes and Prost, 1993), the effective theory of (spin) 
nematics is written in terms of a tensor order parameter, for a three
component spin $\langle q_{\alpha \beta} \rangle = \langle n_{\alpha} n_{\beta} - 1/3 \delta_{\alpha, \beta} \rangle$ (
$\alpha, \beta = x, y, z $), which is clearly invariant under $\vec{n} \rightarrow - \vec{n}$ (identification of
the poles). However, it was only quite recently realized by Lammert, Rokshar and Toner, 1995,
that the complete effective theory should explicitly
incorporate the {\em Ising gauge invariance} associated with the director, which is automatic when the theory
is written in terms of the redundant vector degrees of freedom $\vec{n}$. It is an Ising gauge invariance 
because the vector is defined modulus its sign.
For the spin-nematic, this gauge theory is as follows in Lagrangian formulation.
Define a 3d cubic lattice (2 space and 1 time direction) and define on every site a $O(3)$ vector $\vec{n}_i$.
If these were coupled by normal exchange interactions this would just correspond with the $O(3)$ QNLS. However,
define now the Ising variables $\tau^3_{ij}$ living on the bond between sites $i, j$, taking the values $\pm 1$.
The effective action describing the spin-director order parameter theory is,
\begin{equation}
S = - J \sum_{<ij>} \tau^3_{ij} \vec{n}_i \cdot \vec{n}_j + K \sum_{\Box} \Pi_{\Box} \tau^3_{ij}
\label{z2gauge}
\end{equation}
where the last term is the plaquette action defining Ising gauge theory:
 take the product of the values of the $\tau^3$'s 
living on the bonds of one particular plaquette, 
to sum subsequently over all plaquette's. It is easily checked that
this theory is invariant under the simultaneous local transformations $G_i = \sum_{j(i)} \tau^1_{ij}$ with $\tau^1 | \pm \rangle
= | \mp \rangle$, i.e. reversing all signs of the bonds emerging from site $i$, and the north-south identification
$\vec{n} \rightarrow -\vec{n}$. Hence, this is the famous $Z_2$ lattice gauge theory, minimally coupled to an $O(3)$ matter field.

It is well known (Kogut, 1979) that the $Z_2$ pure gauge theory has a confining  
and a deconfining phase. A representative gauge-fixed configuration
for the deconfining phase is simply the one where all $\tau$'s take the value $+1$. Adding the matter fields an keeping this
configuration of gauge fields, an ordered 
state exist with all vectors pointing in the same direction.  By repeated gauge transformations starting from this gauge-fixed
configuration, one recovers the spin nematic 
order as we just discussed. With a little thought one can convince oneself that the abstract $Z_2$ gauge degree of freedom of
Eq. (\ref{z2gauge}) acquires in this stripe context a material meaning. In the ordered stripe phase, the sublattice parity
has acts like a global symmetry. Having decided that a particular domain has sublattice parity $+1$ one can unambiguously establish
the sublattice parity of any other domain. However, when charge is disordered this is no longer possible. Instead, at energies
less than the charge gap, the sublattice parity survives as a $Z_2$ degree of freedom governed by {\em local} symmetry!

This is not all because two other phases are possible. The $Z_2$ sector can stay deconfining while the $O(3)$ sector disorders
(i.e., keep $K$ large and decrease $J$). This corresponds in the stripe language with a disordering of the hidden spin order,
driven for instance by the frustrations associated with local stripe break-ups, while the sublattice parity order stays intact.
Finally, the $Z_2$ sector can confine. It is well understood that this confining phase corresponds with the ordered phase of
a dual {\em global} Ising model where the dual variables correspond with the topological excitations of the Ising gauge
theory, the `fluxons' or `visons'. These can be visualized in the same gauge fix as for the deconfining phase as lines
cutting through the bonds starting at some point and disappearing to infinity where all $\tau$'s have reversed their sign.
This is like a half-infinite line of antiferromagnetic bonds living in a ferromagnet and one directly recognizes that
this is precise representation of the frustration caused by the {\em stripe dislocations} in the spin
system. Hence, the confining phase of the theory Eq. (\ref{z2gauge}) describes the stripe quantum liquid characterized 
by free dislocations. The material meaning of confinement is that at scales large compared to the characteristic dislocation
distance the notion of sublattice parity no longer exists, even not locally.

Let us now return to the question we posed in the beginning: can the suspected hidden order associated with the high
Tc phenomenon be related to stripes? Obviously, the implicit suggestion throughout this paper has been that sublattice 
parity order is a candidate which should be taken seriously. To be more explicit, could it be that the quantum-criticality
of the optimal Tc superconductors is associated with the zero-temperature transition where the stripe
dislocations unbind? Let us present some arguments favoring this possibility:\\
(i) We have hopefully convinced the reader that sublattice parity order is a genuine part of the vacuum structure of
the stripe phase. At the same time, it has to be that the very notion of sublattice parity is destroyed at 
sufficiently large doping. Although convincing experimental evidence is still missing, one would expect a more
conventional Fermi-liquid/BCS physics if the hole density becomes sufficiently large and sublattice parity does not
exist as a degree of freedom in conventional fermiology. In the above we have spelled out the unique way in which
sublattice parity gets destroyed. Sublattice parity can persist as a degree of freedom governed by local symmetry
even in a state which is spin- and charge wise quantum disordered. This is of course the deconfining state of the
Ising gauge theory. The meaning of the {\em confining} state is that at low energies the whole notion of sublattice
parity has disappeared from the long wavelength theory. Hence, the BCS superconductor is at the same time  a
sublattice parity {\em confining} phase. Since we seem to know the low hole density (stripes) and high hole density
(BCS) limits, it has to be that there is phase transition in between where sublattice parity gets confined. 
A priori, it cannot  be excluded that this phase transition is of the first kind where necessarily the disappearance
of sublattice parity goes hand in hand with other symmetry changes like charge- and/or  spin disordering (e.g.,
the first order transition in the spin nematic where both $\vec{n}$ gets disordered and $Z_2$ confines). However,
this appears as unlikely, given the abundant evidence for quantum-criticality associated with the stripes.\\
(ii) Above all, sublattice parity is a very hidden degree of freedom. There is no existing experiment which can directly
measure if sublattice parity order exists or not, especially so in the charge- and spin disordered phases where it
only exists as a local degree of freedom. It does have indirect consequences which are accessible to experiment at
least in principle. For instance, it can be easily seen that the elementary excitations of the quantum spin-nematic
are associated with $S=2$ instead of the usual triplets (Zaanen, unpublished) and these cannot be directly
measured with neutrons (see also Andreev and Grishchuk, 1984). It could well be that massive excitations of
this kind exist because they are again very hard to measure.\\
(iii) There has been little mention of superconductivity. The reason is that superconductivity comes for free, at
least under the assumption that the stripe phase itself is associated with pairs of electrons. Charge-wise the stripe phase
is then a bosonic crystal and superconductivity cannot be avoided when the charge quantum disorders. Just
as in the case of superconductivity and  antiferromagnetism (Zaanen, 1999; van Duin and Zaanen, 2000)
the mode-couplings between the superconducting phase mode and all other modes are supposed to be irrelevant
operators and therefore superconductivity can be simply superimposed on the physics of the spin- and sublattice
parity sectors, at least at zero temperature.\\
(iv) There is one very serious problem: there is no obvious place for the nodal fermions in this framework. This
is partly by construction. In the above we have assumed that {\em everything} is bosonic and in a bosonic 
universe there is no room for $S=1/2$ excitations. Nodal fermions have surely to do with this spin quantum number.
We already announced that we made crucial assumption by neglecting the  Marshall sign's. As is well known, a $S=1/2$ Heisenberg
spin problem defined on a bipartite, nearest-neighbor bond lattice has a bosonic ground state in the sense that the ground
state wave function is nodeless. The reader familiar with this construction will immediately infer that in the presence
of the stripe dislocations this is no longer true. Hence, at the moment stripes are no longer perfectly connected
the spin system acquires a non-trivial sign structure in vacuum and one no longer knows what to expect (Weng {\em et al.},
1997). Is this
the missing ingredient, linking the above to high Tc superconductivity? We have no clue, but we hope to have convinced
the reader that it is very unreasonable to believe that the relationship between stripes and superconductivity is understood.

Note added in press: after completion of this manuscript we learned that the prediction of Zaanen, 2000, of an
induced modulus string gas which has a stretched-exponential dependence on the density has been confirmed
by numerical simulations (Yoshihiro Nishiyama, preprint, Okayama University).

{\em Acknowledgements.} 
This work has profited from stimulating interactions with numerous theorists and experimentalists
working in the field of high Tc superconductivity and stripes.
A number of theorists who were directly involved in earlier stages of this research deserve
a special mention: Wim van Saarloos, Henk Eskes, Mark Horbach, and Rob Grimberg. The same is true
for Sergei Mukhin and Steve Kivelson because of their large influence on the work in progress.
This research was supported in part  by the Foundation
of Fundamental Research on Matter (FOM), which is sponsored by the
Netherlands Organization of Pure research (NWO), and by the National Science Foundation grant no. PHY99-07949.   

{\bf References}\\

Aeppli, G., Mason, T.E., Hayden, S.M., Mook, H.A., Kulda, J., 1997, Science {\bf 278}, 1432\\
Affleck, I, Halperin, B.I., 1996,  J. Phys. A {\bf 29}, 2627\\
Anderson, P.W., 1997, {\em The Theory of High Tc Superconductivity} (Princeton University Press, Princeton)\\
Andreev, A.F., Grishchuk, I.A., 1984, Sov. Phys. JETP {\bf 60}, 267\\ 
Binder, K.,  1997,   {\em Monte Carlo Simulations in Statistical Physics: an introduction} (Springer Series in solid
state sciences {\bf 80}, Springer, New York)\\ 
Bosch, M., van Saarloos, W., Zaanen, J., 2001, Phys. Rev. B {\bf 63}, 501\\
Carlson, E.W., Orgad, D., Kivelson, S.A., Emery, V.J., 2000, Phys. Rev. B {\bf 62}, 3422\\
Castellani, C., Di Castro C., Grilli, M., 1995, Phys. Rev. Lett. {\bf 75}, 4650\\
Castro-Neto, A.H., Hone, D., 1996, Phys. Rev. Lett. {\bf 76}, 2165\\
Castro-Neto, A.H., Guinea, F., 1998,  Phys. Rev. Lett. {\bf 80}, 4040\\
Chakravarty, S., Halperin, B.I., Nelson, D., 1989, Phys. Rev. B {\bf 39}, 2344\\
Chernyshev, A.L., Castro Neto, A.H., Bishop, A.R., 2000, Phys. Rev. Lett. {\bf 84}, 4922\\
Chubukov, A.V., Sachdev, S., 1993, Phys. Rev. Lett. {\bf 71}, 169\\
Coppersmith, S.N., Fisher, D.S., Halperin, B.I., Lee, P.A., Brinkman, W.F., 1982, Phys. Rev. B {\bf 25}, 349\\
Curro, N.J., Hammel, P.C., Suh, B.J., Huecker, M., Buechner, B., Ammerahl, U., Revcolevschi, 2000, Phys. Rev. Lett. 
{\bf 85}, 642\\\
Dagotto, E., 1994, Rev. Mod. Phys. {\bf 66}, 763\\
Dagotto, E., Rice, T.M., 1996, Science {\bf 271}, 618\\ 
Dai, P.C., Mook, H.A., Hayden, S.M., Aeppli, G., Perring, T.G., Hunt, R.D., Dogan, F., 1999, Science {\bf 284}, 1344\\
De Gennes, P.-G., Prost, J., 1993,  {\em The Physics of Liquid Crystals} (Clarendon Press, Oxford)\\ 
Den Nijs, M., Rommelse, K., 1989,  Phys. Rev. B {\bf 40}, 4709\\
Dimashko, Y.A., Smith, C.M., Hasselmann, N. Caldeira, A.O., 1999, Phys. Rev. B {\bf 60}, 88\\
Du Croo-De Jongh, M.S.L., van Leeuwen, J.M.J., van Saarloos, W., 2000, Phys. Rev. B {\bf 62}, 14844\\
Emery, V.J., Kivelson, S.A., Zachar, O., 1997, Phys. Rev. B {\bf 56}, 6210\\
Emery, V.J., Kivelson, S.A., Tranquada, J., 1999, Proc. Natl. Acad. Sci. USA, {\bf 96}, 8814\\
Emery, V.J., Fradkin, E., Kivelson, S.A., Lubensky, T.C., 2000, Phys. Rev. Lett. {\bf 85}, 2160\\
Eskes, H., Grimberg, R., van Saarloos, W., Zaanen, J., 1996, Phys. Rev. B {\bf 54}, R724\\
Eskes, H., Osman, O.Y., Grimberg, R., van Saarloos, W., Zaanen, J., 1998, Phys. Rev. B {\bf 58}, 6963\\
Evertz, H.G., Lana, G., Marcu, M., 1993, Phys. Rev. Lett. {\bf 70}, 875\\
Fleck, M., Lichtenstein, A.I., Pavarini, E., Oles, A.M., 2000, Phys. Rev. Lett. {\bf 84}, 4962\\
Fradkin, E., 1991, {\em Field theories of condensed matter systems} (Addison-Wesley, Redwood City)\\ 
Haldane, F.D.M., 1981, Phys. Rev. Lett. {\bf 47}, 1840\\
Hasselmann, N., Castro-Neto, A.H.C., Smith, C.M., Dimashko, Y., 1999, Phys. Rev. Lett. {\bf 82}, 2135\\
Helfrich, 1978, Z. Naturforsch. A {\bf 33}, 305\\
Hunt, A.W., Singer, P.M., Thurber, K.W., Imai, T, 1999, Phys. Rev. Lett. {\bf 82}, 4300\\
Kivelson, S.A., Emery, V.J., 1996, Synthetic Met. {\bf 80}, 151\\ 
Kivelson, S.A., Fradkin, E., Emery, V.J., 1998, Nature {\bf 393}, 550\\
Kogut, J.B., 1979, Rev. Mod. Phys. {\bf 51}, 659\\
Klauss, H.H., Wagener, W., Hillberg, M., Kopmann, W., Walf, H., Litterst, F.J., Hucker, M., Buchner, B., 
2000, Phys. Rev. Lett. {\bf 85}, 4590\\
Kruis, H.V., Nussinov, Z., Zaanen, J., 2001, in preparation\\
Lammert, P.E., Rokhsar, D.S., Toner, J., 1995, Phys. Rev. E {\bf 52}, 1778\\
Laughlin, R.B., 1998, Adv. Phys. {\bf 47}, 943\\
Lee, C.H., Yamada, K., Endoh, Y., Shirane, G., Birgeneau, R.J., Kastner, M.A., Greven, M., Kim, Y.J., 2000,
J. Phys. Soc. Jpn. {\bf 69}, 1170\\
Martin, G.B., Gazza, C., Xavier, J.C., Feiguin, A., Dagotto, E., 2000, Phys. Rev. Lett. {\bf 84}, 5844\\
Millis, A.J., Monien, H., 1993, Phys. Rev. Lett. {\bf 70}, 2810\\
Mook, H.A., Dogan, F., 1999, Nature {\bf 401}, 145\\
Mook, H.A., Dai, P.C., Dogan, F., Hunt, R.D., 2000, Nature {\bf 404}, 729\\
Mook, H.A., 2001, private communications\\
Morais-Smith, M., Dimashko, Y.M., Hasselmann, N., Caldeiro, A.O., 1998, Phys. Rev. B {\bf 58}, 453\\
Mukhin, S.I., van Saarloos, W., Zaanen, J., 2001, in preparation.\\
Ogata, M., Shiba, H., 1990, Phys. Rev. B {\bf 41}, 2326\\
Orgad, D., Kivelson, S.A., Calson, E.W., Emery, V.J., Zhou, X.J., Shen, Z.X., 2001, cond-mat/0005457\\
Osman, O.Y., 2000,  thesis, Leiden University\\
Pokrovsky, V.L., Talapov, A.L., 1979, Phys. Rev. Lett. {\bf 42}, 65\\
Polyakov, 1987, {\em Gauge Fields and Strings} (Harwood Ac. Publ., Chur, Switzerland)\\
Pryadko, L.P, Kivelson, S., Hone, D.W., 1998, Phys. Rev. Lett. {\bf 80}, 5651\\
Pryadko, L.P., Kivelson, S.A., Emery, V.J., Bazaliy, Y.B., Demler, E.A., 1999, Phys. Rev. Lett. {\bf 60}, 7541\\
Read, N., Sachdev, S., 1989, Phys. Rev. Lett. {\bf 62}, 1694\\
Sachdev, S., 1999, {\em Quantum Phase Transitions} (Cambridge University Press, Cambridge)\\
Sachdev, S., 2000, Science {\bf 288}, 475\\
Schulz, H.J., 1993, in {\em Correlated Electron Systems}, ed. V.J. Emery (World Scientific, Singapore)\\
Stojkovic, B.P., Yu, Z.G., Chernyshev, A.L., Bishop, A.R., Castro Neto, A.H., Gronbech-Jensen, N., 2000, 
Phys. Rev. B {\bf 62}, 4353\\ 
Tchernyshyov, O., Pryadko, L.P., 2000, Phys. Rev. B {\bf 61}, 12503; Physica C {\bf 341}, 1791\\
Teitelbaum, G. B., Aby-Shiekah, I.M., Bakharev, O., Brom, H.B., Zaanen, J., 2001, Phys. Rev. B {\bf 63}, 020507(R)\\
Tranquada, J.M., Sternlieb, B.J., Axe, J.D., Nakamura, Y., Uchida, S., 1995, Nature {\bf 375}, 561\\
Tranquada, J.M., Ichikawa, N., Uchida, S., 1999, Phys. Rev. B {\bf 59}, 14712\\
Tworzydlo, J., Osman, O.Y., van Duin, C.N.A., Zaanen, J., 1999, Phys. Rev. B {\bf 59}, 115\\
van Beijeren, H., Nolten, I., 1987,  in {\em Structure and Dynamics of Surfaces II}, eds. W. Schommers and
P. von Blanckenhagen (Springer, Berlin)\\
van Duin, C.N.A.,  Zaanen, J., 1997, Phys. Rev. Lett. {\bf 78}, 3019\\
van Duin, C.N.A.,  Zaanen, J., 1998, Phys. Rev. Lett. {\bf 80}, 1513\\
van Duin, C.N.A.,  Zaanen, J., 2000, Phys. Rev. B {\bf 61}, 3676\\
Voit, J., 1994, Rep. Prog. Phys. {\bf 57}, 977\\
Voita, M., Zhang, Y., Sachdev, S., 2000, Phys. Rev. B {\bf 62}, 2000\\
Voita, M., Sachdev, S., 1999, Phys. Rev. Lett. {\bf 83}, 3916\\
Weng, Z.Y., Sheng, D.N., Chen, Y.C., Ting, C.S., 1997, Phys. Rev. B {\bf 55}, 3894\\
White, S.R., Scalapino, D.J., 1998, Phys. Rev. Lett. {\bf 81}, 3227; Phys. Rev. Lett. {\bf 80}, 1272\\
Wiese, U.J., Ying, H.-P., 1994, Z. Phys. B {\bf 93}, 147\\ 
Yamada, K., Lee, C.H., Kurahashi, K., Wada, J., Wakimoto, S., Ueki, S., Kimura, H., Endoh, Y., 
Hosoya, S., Shirane, G., Birgeneau, R.J., Greven, M., Kastner, M.A., Kim, Y.J., 1998, Phys. Rev. B {\bf 57}, 6165\\
Zaanen, J., Gunnarsson, O., 1989, Phys. Rev. B {\bf 40}, 7391\\
Zaanen, J., Horbach, M.L., van Saarloos, W., 1996, Phys. Rev. B {\bf 53}, 8671\\
Zaanen, J., Osman, O.Y., Eskes, H., van Saarloos, W., 1996, L. Low Temp. Phys. {\bf 105}, 569\\
Zaanen, J., van Saarloos, W., 1997, Physica C {\bf 282}, 178\\
Zaanen, J., 1998, J. Phys. Chem. Solids {\bf 59}, 1769\\
Zaanen, J., Osman, O.Y., van Saarloos, W., 1998, Phys. Rev. B {\bf 58}, R11868\\
Zaanen, J., 1999, Physica C {\bf 318}, 217\\
Zaanen, J., 1999, Science {\bf 286}, 251\\
Zaanen, J., 2000, Phys. Rev. Lett. {\bf 84}, 753\\
Zaanen, J., 2000, Nature {\bf 404}, 714\\
Zaanen, J., Nussinov, Z., 2001, Physica C (in press, cond-mat/0006193)\\
Zachar, O., 2000, Phys. Rev. B {\bf 62}, 13836\\

\end{document}